\documentstyle[aps,epsf]{revtex}
\begin{document}
 \newcommand \be {\begin{equation}}
\newcommand \ee {\end{equation}}
 \newcommand \bea {\begin{eqnarray}}
\newcommand \eea {\end{eqnarray}}
\newcommand \nn \nonumber
\def \vt{\hat{{\bf{t}}}}
\def \vn{\hat{{\bf{n}}}}
\def \vu{\hat{{\bf{u}}}}
\def \va{{\bf{a}}}
\def \vb{{\bf{b}}}
\def \ve{\hat{{\bf{e}}}}
\def \vz{\hat{{\bf{z}}}}
\def \vx{\hat{{\bf{x}}}}
\def \vy{\hat{{\bf{y}}}}
\def \vr{{\bf{r}}}
\def \vw{\hat{{\bf{w}}}}
\def \vm{\hat{{\bf{m}}}}
\def \vX{{\bf{X}}}
\def \vY{{\bf{Y}}}
\def \vZ{{\bf{Z}}}
\def \vF{{\bf{F}}}
\def \vA{{\bf{A}}}
\def \vB{{\bf{B}}}
\def \vS{{\bf{S}}}
\def \vl{{\bf{l}}}
\def \vp{{\bf{p}}}
\def \Om{{\bf{\Omega}}}
\def \grad{{\bf{\nabla}}}
\def \vnu{{\bf{\nu}}}
\def \Up{{\bf{\Upsilon}}}
\def \(({\left(}
\def \)){\right)}
\def \eps{\epsilon}
\title{\begin{flushright}
{\rm LPTENS  /99/04 \\
cond-mat /9904018 \\
December 1999 }
\end{flushright}
Elastic Rod Model of 
a Supercoiled DNA Molecule}
\author{C. Bouchiat$^{1}$ and M. M\'ezard$^{1,2}$}
\address{$^1$Laboratoire de Physique Th\'eorique de l'Ecole Normale Sup\'erieure
\footnote{UMR 8548:  Unit\'e Mixte du Centre National de la Recherche 
Scientifique \\ 
 et de l'\'Ecole Normale Sup\'erieure. } 
\\
24 rue Lhomond 75231 Paris Cedex 05,  France \\
$^2$
Institute for Theoretical Physics, 
University of California Santa Barbara, CA 93106-4030, USA.
}

\maketitle

\begin{abstract}
    We study the elastic behaviour of a supercoiled DNA molecule.
    The simplest model is that of a rod like chain, involving two elastic
    constants, the 
    bending and the twist rigidities.
Writing this model in terms of Euler angles, 
 we show that the corresponding Hamiltonian is singular
    and needs a small distance cutoff, which is a natural length scale giving
    the limit of validity of the model, of the order of the double helix 
    pitch. The rod like chain in presence of the cutoff is able to reproduce
    quantitatively the experimentally observed effects of supercoiling
    on the elongation-force characteristics, in
    the small supercoiling regime. An exact solution of the model,
    using both transfer matrix techniques and its mapping to a quantum mechanics
    problem, allows to extract, from the experimental data, the value of the twist rigidity.
    We also analyse the variation of the torque and the writhe to twist ratio
    versus supercoiling, showing analytically the existence of a rather sharp
    crossover regime which can be related to the excitation of plectoneme-like 
    structures. Finally we study  the extension fluctuations   of a stretched
    and supercoiled DNA molecule, both at fixed torque and at fixed
    supercoiling angle, and we compare the theoretical 
    predictions to some preliminary experimental data.

\end{abstract}

\section{Introduction}
During the last few years, Single Molecule Biophysics has
become a very active field of research. Among the new 
 explored  areas, one finds the study of elastic properties of 
biopolymers, in particular the DNA molecule, under physical
conditions close to that encountered in living organisms 
\cite{smith,perk,strick}. 
The theoretical work reported in the present paper was
motivated by the results of micromanipulation experiments being 
performed at Ecole Normale Sup\'erieure \cite{strick} with
an isolated DNA molecule, immersed in a solution of given 
salinity at room temperature. One extremity of the molecule
is  biochemically bound at multiple sites to a treated glass cover
slip.
The  other end is similarly attached to a 
paramagnetic bead with a radius of a few microns
which is  performing a Brownian  motion in the 
solution. An appropriate magnetic device is able to 
pull and rotate at the same time the magnetic bead.
Because of the multiple attachments of the  molecule 
 extremities, the rotation of the bead
results in  a supercoiling constraint for the DNA isolated
molecule.

When no rotation is applied on the bead, the   force versus extension curves 
are very well described \cite{marsig}
 by a simple elastic model, the  so-called Worm Like Chain (WLC) model
 which
describes a phantom chain  using a  
  single elastic constant, the bending rigidity \cite{fixman}. This model
 reproduces the data within experimental accuracy, in
  a very wide range of pulling force from
$F \simeq 0.01\, pN$ to $F\simeq 10\, pN$. One has  to go up to $70\, pN$
to leave the elastic regime and see a sharp increase of the
molecule length which  is associated  with a transition to a new
molecular phase, the "S-DNA" \cite{sDNA}. 

In contrast, in stretching experiments
performed with supercoiled   DNA  molecules \cite{strick,strick2},
 two molecular structural
transitions appear  within a more restricted domain of forces.  
An instructive  way to display the   supercoiled  DNA 
stretching  data is to plot the molecule extension  versus
the  degree of supercoiling $\sigma$ (defined as the ratio of the number of
 bead turns  to the number of 
double  helix turns in the relaxed molecule), at various fixed
values  of the stretching force. Three such plots
are shown on  Figure \ref{fig1} for typical force values.
 Working with molecules having   a moderate degree of  supercoiling,
$\sigma \le 0.1$,
 two changes of regime occur in the force range between
$F=0.01\, pN $ and $ F=5.0\, pN $. For forces below $  0.44\, pN $ 
the extension versus  supercoiling curves are symmetric under
 the exchange $ \sigma \rightarrow - \sigma  $
 and have been called `hat' curves for obvious reasons. 
 This is the domain of entropic elasticity 
which will be our main topic in the present paper. When $ F \ge 0.5\, pN $ 
a plateau is developed in the underwinding region $ \sigma <0 $. The
mechanical
underwinding energy is transformed into  chemical energy which is used to
open
up the hydrogen bonds and denaturation bubbles appear along the DNA chain.
Above $4 \,pN$, a plateau appears also in the overwinding region  $ \sigma >0 $.
This has been interpreted as a transition to a new structure of DNA called 
the " DNA-P"\cite{pDNA}. The double helix structure is clearly responsible  for
the ability of  DNA to convert mechanical torsional energy into chemical 
energy, with a  net  preference  for underwinding energy. 
  
  Our work here will be restricted to the
 entropic elastic regime, and its aim
  is primarily to compute the extension versus supercoiling hat curves.
 We have used the simplest extension of the WLC model,
 which is still a phantom chain,
with one single new elastic constant, the twist rigidity.
 We call the corresponding model a  Rod Like Chain  (RLC). 
  We will show that the RLC model,
 described in terms of local Euler angle variables,
 is singular in the limit of
  a purely continuous chain. Its physical realization requires the introduction
  of a short length scale cutoff (which will turn out to be of the order
  of magnitude of the double helix pitch). Then the RLC is able to
  reproduce all experimental data in the   restricted domain
  of its validity which we have discussed above. The good fit to 
  the experimental data allows to deduce 
the  twist rigidity to bending rigidity ratio, 
  which was rather poorly  known so far (with uncertainties
  up to a factor or 2).
  A short account of some parts of this work has been discussed
  in a previous letter \cite{BM1}
(where the RLC model appeared under a slightly different name: the WLRC
model).
 We shall give here a more comprehensive - and hopefully comprehensible- description
of our model. Our principal  aim is to show how the RLC model, once it has been 
properly formulated, can be solved exactly: by exact, we mean that the 
predictions of the model can be computed  by  a mixture of analytic and
numerical methods  to any desired precision.

The first conclusive work able to fit the experimental data from
a rod like chain with two elastic constants is that of
Marko and Vologodskii \cite{marvol}, who used a discretized model
and performed some Monte Carlo simulations. On the analytical side,
Fain et al. \cite{fain} wrote the elastic energy in terms of a local
 writhe formula,
similarly to the one which we have used, but they used it only in the 
zero temperature limit. In an independent work which appeared just after ours,
Moroz and Nelson \cite{mornel1} used the same form of energy as we did,
but with a different approach for handling  the singularity of
the model in the continuum limit. We shall comment extensively on the
differences of these approaches, as well as on the obtained results.
 
Developing a theory beyond the elastic regime is a much more difficult task,
since such a theory should involve both the self repulsion of the chain
(necessary to prevent a collapse of plectonemes), and the possibility
of denaturation. Some first attempts in these directions can be found
in \cite{liverpool,zhou}.

Recently an  explicit model, coupling 
the hydrogen-bond opening with the untwisting of the 
double helix, has been proposed \cite{cocco}. It leads to a 
unified description of thermal and untwisting DNA denaturation
in the high stretching force limit.

We  now  indicate  the organization of the paper,
with some  emphasis upon the  "RLC model crisis" and its solution.
In sect. \ref{el_en_RLC} we introduce
the RLC model and discuss its range of validity. We note first  that
the  elastic energy used in the present paper is invariant under rotation
about the molecular axis, in apparent contradiction with the double helix structure.
 We  prove that added  cylindrical  asymmetric  terms in the elastic energy  
are washed out by  averaging upon the empirical length resolution $ \Delta l $, 
  about  three times the double helix pitch $p$.  
 We also discuss the   relationship to
the Quantum Mechanics   problem of a symmetric top, which appears naturally 
when one considers the configurations of the DNA chain as world lines
in a quantum mechanical problem. We stress that, despite a formal analogy,
there is an important difference in the formulation  of the two problems.
The RLC analog of the  angular momentum is not quantized since,   
in  contrast to the  symmetric top, the physical states of an  elastic rod are
 not invariant  under rotations of $ 2 \pi$   about its axis. This is
 the origin of  the "RLC crisis".

In Sect. \ref{part_funct_DNA} we incorporate in the   RLC  partition function    
 the supercoiling constraint coming from  the rotation of the magnetic bead. 
First written as a boundary condition upon Euler angles used to specify 
the DNA configuration, the supercoiling constraint is transformed,
assuming some regularity conditions on the Euler angles, 
into an equality between the bead rotation angle  and the sum  of the 
twist and writhe variables of the open molecular chain. These variables
fluctuate independently in absence of supercoiling so that the constrained 
RLC partition function  is given as a convolution
product of the twist and writhe partition functions. The writhe 
partition function Fourier transform is in correspondence  with the Quantum 
  Mechanics problem of  a charged particle moving in the field of
a magnetic monopole with an unquantized charge.
The spontaneous fluctuations of the writhe provide a spectacular
signature of the RLC model pathology: the second moment 
of the fluctuation is predicted
to be infinite in the continuum limit. 
This divergence comes from a singularity appearing 
in the RLC potential
when the molecular axis is antiparallel to the stretching force.

 Sect. \ref{discrete_theory} explains how the angular cutoff needed
in order to transform the RLC model into a sensible one
can be generated by the discretization of the
chain. The introduction of a length cutoff $b$, of the order $ \Delta l$,
appears  rather natural since an average upon $ \Delta l$ has been invoked 
to justify our choice of the RLC elastic energy. We proceed next
with the development of the  two main tools used  to get an exact 
solution of the discretized RLC model: a direct  transfer matrix approach
and a Quantum Mechanics  approach involving a
 regularized  RLC Hamiltonian  with a potential derived from 
the discretized model and now free from any singularity. 

  In sect. \ref{quant_ana} we use the two  above methods in order to compute the
elongation versus supercoiling characteristics of a long DNA chain. The results are
compared with experimental data in sect. \ref{ana_exp} where the small 
distance cutoff $ b $ and the twist elastic constant are determined. 

Sect. 
\ref{Monte_Carlo} presents a complementary approach, that of
 Monte Carlo simulations which  provides the only method so far
 to introduce the constraints of self-avoidance and knots elimination. 
After a  presentation of the results of  Marko and Vologodskii \cite{marvol},
 we describe our less ambitious simulations which  were used 
to validate our analytic method.

In sect. \ref{tw_wr_pl} we use the RLC model to analyse, 
at a force  taken on the high side
of our explored area, the variation of the torque and the writhe to twist ratio
 versus supercoiling. The two curves exhibit a rather sharp change 
 of regime near one given value of the  supercoiling
angle: the torque, after a nearly linear increase, 
becomes almost supercoiling independent while the writhe to twist ratio,
 initialy
confined to the $ 20 \% $ level, develops a fast linear increase. We give arguments 
suggesting that this quasi-transition is associated with the creation of 
plectoneme-like configurations, able to absorb supercoiling at constant torque.

 Finally the sect. \ref{fluctuations}  studies,
 within the RLC model,  the extension fluctuations   of an isolated  DNA molecule,
  subject both
 to stretching and supercoiling constraints. The predictions are made
for  two thermodynamic ensembles: one at fixed torque, the other at fixed
supercoiling angle. Although the two ensembles lead, as they should, 
to identical results as
far as average values are concerned, they yield widely different 
predictions  for the extension fluctuations above a certain supercoiling threshold
(close  to the one appearing in the previous section): 
the fixed torque ensemble becomes much
more noisy. We give a qualitative explanation for this peculiar behaviour and compare
our predictions to some preliminary experimental data 
at fixed supercoiling.   

 In sect. \ref{conclusion} we give  a summary of the work and some 
 perspectives on its
 future extensions.

\section{Elastic Energy of The Rod-Like Chain}
\label{el_en_RLC}
%
 A given configuration of the rod-like chain (RLC)   is specified, in the 
continuous limit,
 by  the  local orthonormal trihedron  $\{ {\ve}_{i}(s)\}=  \{ \vu(s),\vn(s),\vt(s)  \}$
where $s$ is the arc length along the molecule, $ \vt $ is the unit
vector tangent to the chain, $\vu(s) $ is along the basis line
 and $\vn(s)=\vt(s)\wedge\vu(s)$. The evolution of the trihedron
 $\{ {\ve}_{i}(s)\}$  along the chain 
is obtained by applying a rotation $ {\cal R }(s) $ to a reference trihedron
$\{ {\ve}_{i}^0(s)\}$ attached  to  a rectilinear relaxed molecule:
 ${\ve}_{i}(s) ={\cal R }(s)\cdot  {\ve}_{i}^0(s)$. 

The rotation  $ {\cal R }(s) $ is parameterized
 by  the usual three  Euler angles $\theta(s)$, $\phi(s)$ and $\psi(s)$.
 The  reference trihedron  is such that $\theta(s)=0$,
$\phi(s)+\psi(s)=\omega_{0} s$, where
$ \omega_{0}$ is the rotation per unit length of the base axis in
the relaxed rectilinear DNA molecule.
 With the above  definition, the set of $s$ dependent Euler angles
 $ \theta(s), \phi(s),\psi(s) $ describes the general deformations
 of the DNA molecule with respect to the relaxed rectilinear configuration.
It is now convenient to introduce the angular velocity vector $\Om(s)$ 
associated with the 
rotation $ {\cal R }(s)$. The evolution of the trihedron  $\{ {\ve}_{i}(s)\}$ 
along the molecular
 chain is given in term of $\Om(s)$ by the set of equations:
\be  
 {d\,\over ds }\, {\ve}_{i}(s) =\(( \Om(s) + \omega_{0} {\vt}(s) \))\wedge 
{\ve}_{i}(s) 
 \ee
 We shall use in the following 
 the components of $\Om(s) $ along the  trihedron  $\{ {\ve}_{i}(s) \}$:
 $ {\Omega}_{i}= \Om(s) \cdot {\ve}_{i}(s)  $ with $  {\ve}_{1}=\vu,\; 
{\ve}_{2}=\vn $ and 
 $ {\ve}_{3}=\vt$. ( The $ {\Omega}_{i} $ are computed  in terms of the Euler 
angles
 and their $ s $ derivatives in the Appendix A).
The stretched RLC  energy $E_{RLC}$ is written as a line integral  along the 
chain of length $
L$,  involving a sum  of three contributions: 
\be
 E_{RLC}= k_{B} T\,\int_0^{L} \, ds \left( {\cal E}_{bend}(s)+{\cal 
E}_{twist}(s)
+{\cal E}_{stretch}(s) \right)
\label{EL0}
\ee 
The above  bend, twist and stretch linear  energy densities are given by:    

\bea 
{\cal E}_{bend} &=&  {A \over 2}\,\left( {\Omega}_{1}^2+{\Omega}_{2}^2 \right)=
                 {A \over 2}\, \vert  {d \vt(s) \over ds} \vert ^2 = 
 {A \over 2} \left(
{\dot{\phi}}^2 \,{\sin^2\theta}+{ \dot{\theta}}^2 \right)  \nn \\
{\cal E}_{twist} &=& {C \over2}\,{\Omega}_{3}^2=
 { C \over 2} \,( \dot{\psi} +\dot{\phi} \, \cos\theta )^2  \nn   \\ 
{\cal E}_{stretch} &=&  - \vt(s)\cdot\vF /( k_{B} T)= - {F  \cos \theta \over k_{B} 
T}
\label{EL}
\eea
where  the dot stands for the  $s$ derivative.

$ {\cal E}_{bend} $ is proportional to the inverse of 
	the square of the curvature radius and  represents the resistance 
against bending around
 an axis perpendicular to the chain axis. The coefficient $ A $ is 
  the persistence length; its  typical value is $ 50 
nm$.

${\cal E}_{twist}$ is proportional to the square of the
 projection of the angular velocity along the 
molecular axis and gives the energy associated with a twist constraint.
 The
twist rigidity $C$ is known to be in the range 50 to 100 nm; its determination from the
experimental measurement will be the topic of section \ref{ana_exp}.

${\cal E}_{strech}$ is the potential energy 
associated with the uniform stretching force  $ \vF=F.\vz $ which is applied at 
the free end of the molecule; it is written as a line integral involving
 $ \vt(s).\vz $.

The ground state configurations of the rod like chain have been studied
in \cite{fain}. These may apply to the properties of small plasmids, but
in the regime which we are interested in, the thermal fluctuation effects
are crucial.
 If no twist dependent properties  are imposed or measured,
  one can factor out the $\psi$ integral in the partition function,
  or equivalently use the  $ C=0$ theory, in which case one recovers
the elastic energy of the Worm Like Chain model :
\be 
{E_{WLC} \over k_{B} T}= \int_0^{L} \, ds \((   {A \over 2}\, \vert  {d \vt(s) 
\over ds} \vert ^2
 -\cos\theta(s) F   /( k_{B} T) \))
\label{EWLC}
\ee 
It is now appropriate to raise two questions  about the above formulas.
\begin{center}  
\subsection{ Is a cylindrical  symmetric elastic  rigidity tensor adequate for 
the DNA chain? }
 \end{center}
 The theory which we are using here is an elastic theory which cannot be
 valid down to atomic scales. In particular it does not take into 
 account the microscopic charges on the DNA and
 their Coulomb interactions  in the solvent, but it just describes the net effect
 by a set of elastic constants. However, even at the level of an elastic description,
 one may wonder whether our choice of elastic tensor, which 
 ignores the helical structure of the DNA, is valid. Indeed 
 the elastic energy given by equations (\ref{EL}) 
involves a 
rigidity  tensor with a cylindrical  symmetry around the molecular axis $ 
\vt(s) $. 
 We would like to argue that such a description is a reasonable approximation
  if one deals with experimental data
obtained  with a finite  length resolution $ \Delta \, l
 \simeq 10 nm$, as is the case for the
experiments we shall analyse.
 The pitch $ p$ of the double helix contains 
approximately 10 basis which corresponds
to a length   of $ 3.4 nm$. A simple way to break the cylindrical  symmetry is 
to introduce
in the elastic energy linear density a term proportional to 
$ \Delta \Omega(s)=  {\Omega}_{1}^2-{\Omega}_{2}^2 $.
As it is shown in Appendix A, $ \Delta\Omega(s) $ can be written in terms of the 
Euler angles
in the following way: 
$$
\Delta\Omega(s) = {\Omega_{\perp}}^2 \,\cos 2( \zeta(s) +\psi(s)+ \frac{2 \pi 
s}{p})
$$
 with $ {\Omega_{\perp}}^2  ={\Omega}_{1}^2+{\Omega}_{2}^2 $ and 
$\zeta(s) =\arctan\((\dot{\theta}/(\sin\theta\dot{\phi} )\))$;

One must note   that $ \Delta\Omega(s) $ oscillates  with a wavelength which is 
half the pitch $p$ 
of the double helix. An average involving a resolution length, about 6 times 
the oscillation wavelength, is expected to lead to a strong suppression  of $ 
\Delta \Omega(s) $.
The  average $ \overline{  \Delta\Omega(s) }=  \int ds_1 
 P(s_1  -s)\Delta\Omega(s_1)$, associated with the resolution  function
  $ P(s)={1\over \sqrt{2 \pi } \Delta\,l}\exp( -{1\over2} s^2 /{\Delta\,l}^2) $,
 is  computed explicitly  in Appendix A in terms of simple Gauss integrals, assuming  
 that the phase $\zeta(s) +\psi(s)$ and the amplitude ${\Omega_{\perp}}^2$ vary 
linearly within the interval  $ ( s-\Delta\,l, s+\Delta\,l)$.   Ignoring first 
the $\Omega_{\perp} $ variation and  assuming that the supercoiling angle 
per unit length  is $ \ll \omega_0$,   
 we arrive  in Appendix A to  the result: 
$$
\overline{  \Delta\Omega(s) }\approx \Delta\Omega(s) 
\exp\((  -\frac{1}{2} (\frac{4\pi \,\Delta\,l}{p})^2 \))
$$
Taking $ p=3.4 nm $, $\Delta\,l=10 nm $  one gets: $\frac{4\pi \,\Delta\,l}{p} \approx 37$;
it is clear the above expression of $\overline{  \Delta\Omega(s) }/\Delta\Omega(s)$
is zero for all practical purposes. The extra term
 $ \propto { \partial \, {\Omega_{\perp}}^2  \over  \partial \,s}$ is also
found to be exceedingly small.
The same suppression factor holds for $ \overline{ {\Omega}_{1}\,{\Omega}_{2}} $ 
 while  
in the case of   $\overline{ {\Omega}_{1}\,{\Omega}_{3} } $ the argument of  the 
exponential
is divided by 4 but with the value quoted above  for $\Delta\,l$   the 
suppression
effect is still very important.
Therefore  all the  cylindrical asymmetric terms in
 the elastic rigidity tensor are washed out by the empirical  finite length 
resolution averaging.

In the same spirit there are various effects which are not
taken into account by our description. An obvious one is the heterogeneity
of the sequence: a more microscopic description should include some fluctuations of
the rigidities along the chain, depending on the sequence of bases. However one expects
that such fluctuations will be averaged out on some long 
length scales, such as those involved in the experimental situation
under study. This has been confirmed by some more detailed study involving
some simple model of disordered sequences \cite{BDM}.
One should also notice that we have not allowed the total length of the chain to vary.
It is possible to add some elastic energy to bond stretching as well as some
stretch-twist  coupling in order to try to describe the behaviour 
at rather large forces and supercoiling \cite{marko,KLNO,mornel1}. 
But as we will see such terms are 
irrelevant for the elastic regime which we study here.

 
\subsection{ What is the precise connection of the elastic   rod   
thermal fluctuations  problem  with  the  quantum theory of a symmetric top ?}
 
  A careful inspection of the formulas (\ref{EL0}) and (\ref{EL})
 suggests a close analogy of 
the  elastic  linear energy density with the classical lagrangian of the 
symmetric top with A and C being, up to a constant factor, the moments of 
inertia.
For given values of the Euler angles  $ \theta(s), \phi(s),\psi(s) $  
at the two ends $s_0=0$ and  $ s_1=L$ of the chain, the
partition function of the RLC model is given by the 
following path integral:
 \begin{equation}   
 Z({\theta}_1,{\phi}_1,{\psi}_1,s_1\vert{\theta}_0,{\phi}_0,{\psi}_0,s_0 )=
\int {\cal D} \((  \theta, \phi,\psi \)) 
 \exp\((-{E_{RLC}\over k_B T }\))
\label{Zpath}
 \end{equation} 
where $ {\cal D} \((  \theta, \phi,\psi \)) $ stands for the integration 
functional  
measure for  the set of paths  joining  two points of the Euler angles space
with  $ 0 \le \theta \le \pi $ but no constraints  imposed upon $ \phi $ and $ 
\psi $ . 
Let us perform in the elastic energy   integral  over $s$  ( Eq \ref{EL0} ) 
an analytic continuation 
towards 
the imaginary $s$ axis. We shall use  for convenience  a system of units where 
  $ \hbar= c  =k_BT=1 $ and  write $ \Im{s}=t $. The elastic energy  $ E_{RLC} $ is  
transformed into
$-i \int_{t_0}^{t_1} dt {\cal L} (t)$, where the Lagrangian is
that of a symmetric top in a static electric field $f$:
$ {\cal L} (t)=  {1\over2}\sum_1^3 {C}_i {\Omega}_{i}^2 +f \cos 
\theta(t) $, with inertia moments
 $ {C}_1={C}_2= A $, $ C_3=C$ and $ f= F/(k_BT) $.
The analytically  continued   partition 
function
 of the RLC model  is then identified with the Feynmann amplitude:
\bea 
\langle {\theta}_1,{\phi}_1,{\psi}_1,t_1\vert{\theta}_0,{\phi}_0,{\psi}_0 ,t_0 
\rangle &= &
\int {\cal D}\((  \theta,\phi,\psi \)) 
 \exp\((i  \int_{t_0}^{t_1} dt {\cal L} (t) \)) \nonumber \\
&=&
\langle {\theta}_1,{\phi}_1,{\psi}_1\vert\exp \(( -i(t_1-t_0){\hat{\cal 
H}}_{top} \))
 \vert{\theta}_0,{\phi}_0,{\psi}_0 \rangle
\nonumber
\eea
where $ {\hat{\cal H} }_{top} $ is the symmetric top Hamiltonian
 written as a second order differential operator acting  
upon Euler angles wave functions.  
Returning to  real value of s, ones gets the RLC model partition function as
a Quantum Mechanics matrix element, with the substitution $i(t_1-t_0) \to s_1-s_0$.
It is now convenient to introduce the complete set of energy eigenstates
 of ${\hat{\cal H}}_{top}$:
$  {\hat{\cal H}}_{top} \mid n \rangle =E_{n} \mid n \rangle $. One gets
then:
\bea
 Z({\theta}_1,{\phi}_1,{\psi}_1,s_1\vert{\theta}_0,{\phi}_0,{\psi}_0,s_0  ) & = 
&
   \sum_n \,\exp- L\,E_{n} \nonumber \\ 
 & &  {\Psi}_n({\theta}_1,{\phi}_1,{\psi}_1) 
{\Psi}_n^{\star}({\theta}_0,{\phi}_0,{\psi}_0)
\label{ZQM}
\eea 
where $ {\Psi}_{n}(\theta,\phi,\psi) =\langle \theta,\phi,\psi  \vert n  \rangle 
$
is the eigenfunction relative to the state $\vert n \rangle $. 
As we shall discuss later, the interest of the above formula lies
in the fact that the above sum  is dominated by  the term of lowest energy.

 At this point one may get the impression that the solution of the RLC model
is reduced to a relatively straightforward problem of Quantum  
Mechanics. Although  this identification is valid   in the case
of the worm-like-chain model (WLC) which describes the entropic elasticity
 of a stretched  DNA molecule with no twist constraints,  it is not the case for
the more general RLC model. The above analysis was somewhat formal in the sense
 that  the  Hamiltonian $ {\hat{\cal H}}_{top} $ was given as a differential 
operator
  and it is thus not completely defined, until the functional
 space on which it is acting is properly specified.
 Physical considerations will lead us to choose different 
functional
spaces for the two problems in hands:

  { \it In the symmetric top problem the space is that of $2 \pi$ periodic   
functions of the  Euler angles  $ \phi$ and $ \psi  $ , but in
the study of RLC thermal fluctuations the space is that of
 general  functions  of  these angles, without any constraint of periodicity.}

As an illustration, let us discuss the simple case of a non-flexible rod with 
$E={C\over 2} k_{B} T\,\int_0^{L} \, ds  \,{\dot{\psi}(s)}^2$. 
The associated quantum problem is the cylindrical rotator 
described by the Hamiltonian:
$ {\hat{\cal H}}_{ rot}= -{1\over 2 C}\, \frac{ {\partial}^2 }{ {\partial\psi}^2 
} $.
 The $ 2 \pi  $ periodicity  of the motion implies 
 a discrete spectrum for the conjugate momentum
$ p_{\psi}= -i  \frac{ {\partial} }{ {\partial\psi} } $; the corresponding
 wave function is $\exp(i\,k \, \psi )$, where $k$ is an arbitrary integer. 
 Imposing  on the molecular chain the boundary
 conditions $ \psi(s_0=0)=0 \, , \psi(s_1=L)=\chi $, 
the partition function of the non-flexible rod is obtained by a straightforward
application of formula (\ref{ZQM}):
$$ 
Z(\chi,L)= \sum_k \exp\(( i\,k \, \chi -{ L\over 2 C}\, k^2 \)) 
$$ 
In the experiments  to be discussed in the present paper $ L \gg C $
 and, as a consequence, the sum over
 the integer $k$  is dominated by  the terms
 $ k=0,\pm 1$, leading to
 the rather unphysical result: $  Z(\chi,L)=1 +2 \cos \chi \exp(- { L\over 2C}) $.

On the contrary, in order to describe
 the twisting of an elastic rod.
The spectrum of $ p_{\psi} $ has to be taken as continuous 
 and the dicrete sum $ \sum_k ... $ must be replaced
 by the integral $\int_{-\infty}^{\infty} dk/( 2\pi)... $, which  yields
 the correct  result for the partition function:  
\be
Z(\chi,L) \propto  \exp-\(( { C \over 2\, L}\, {\chi}^2 \)) 
\label{ZT}                           
\ee  
 Two significant physical quantities can be easily obtained from the above
partition function: 
First,  the second   moment $ < {\chi}^2 >$ relative to a situation where
 $\chi$ is no longer constrained  but  allowed to fluctuate freely.
  Keeping in mind that $ P( \chi)  \propto Z(\chi,L) $,
one gets : $< {\chi}^2 >= {L\over C}$.
Second,  the torque  $ \Gamma $  given by   the logarithmic derivative 
 of $ Z(\chi,L) $ with respect to $ \chi $:
 $ \Gamma = k_B T  { C \, \chi \over L}$. 
 The above quantities look very reasonable from a physical point of view:

 $ < {\chi}^2 >$ scales linearly in $ L $, as expected from the fluctuations
of a linear chain with a finite correlation length $C $ and the expression
 of $ \Gamma $ reproduces an elasticity textbook formula, if one remembers
that $ k_B T C $ is the usual twist rigidity.

The above considerations may look somewhat simple-minded.
 It turns out   however that they  have  important 
physical consequences.  In order to solve the RLC model, we
shall have to deal with a  quantum spherical top when its angular 
momentum along the top axis is not quantized. Such a problem is 
 mathematically singular and
 as a consequence the continuous limit of the RLC model we have considered  
 so far will not give an adequate description of supercoiled DNA. We shall have
to introduce a discretized version of the RLC model,
 involving an elementary length scale $b$ about twice the double helix 
pitch $p$. This is consistent with the considerations of the previous 
subsection where the  empirical length resolution  $ \Delta\,l \approx 3 p $ was 
invoked in order to justify the cylindrical symmetry of the tensor of elastic 
 rigidities. 
\begin{center}  
 \section{ The Partition Function for Supercoiled DNA in the Rod-like
chain Model }
\label{part_funct_DNA}
\end{center}

 
\subsection{ Implementing the experimental supercoiling constraint on
the partition function }
 The first step is to incorporate 
in the functional integral the  supercoiling  constraint
which results from a   rotation around the stretching  force $ \vF $
of  a    magnetic bead biochemically  glued to the free end of the DNA 
 chain. The  rotation ${\cal R}$ in terms of
 the Euler angles is usually   written as a product of three elementary 
rotations,
 a rotation  $\theta $  about the $y$ axis 
sandwiched between two rotations $ \psi$ and $ \phi$ about the 
$z$ axis, performed in that  order:
${\cal R}= R(\vz,\phi)\,R(\vy ,\theta)\,R(\vz,\psi)$, where $R(\vm,\gamma)$ 
stands for a rotation $\gamma$ about the  unit vector $\vm  $.
  It is convenient, in the present context, to introduce a different form
 of  ${\cal R}$\, 
 involving a product of two rotations written 
in terms of a new set of variables, $\theta ,\psi $ and $ \chi=\phi+\psi $: 
$ {\cal R} =R(\vz,\chi)\,R(\vw(\psi),\theta) $ where  
the  unit vector $\vw(\psi)$   lies in the  $x \; y $ plane and 
 is  given by: $ \vw(\psi)=R(\vz,-\psi) \vy $.
( See Appendix A for a proof of the identity of the two above forms of ${\cal R}$).
Let us consider, before any application  of an external rotation,
the   DNA segment sticking out from the bead. 
Assume that it is short enough 
so that its direction is not affected by thermal fluctuations.
Its orientation is specified by three Euler angles,
  $ { \theta}_{in},{\psi}_{in},{\chi}_{in} $.
 If  the  rotation of the magnetic  bead  by $ n $ turns is performed 
adiabatically, the final orientation of the molecular end trihedron   $\{ 
{\ve}_{i}(L)\} $
  is specified by the rotation:
$$
 {\cal R}(L)= R \(( \vz, 2\pi\,n +{\chi}_{in} \)) \,
 R \(( \vw({\psi}_{in}), {\theta}_{in}\)).   
$$ 
 We can read off   easily  from the above formula: $ 
\chi(L)=\phi(L)+\psi(L)=2\pi\,n +{\chi}_{in} $.
 It is reasonable to assume  that in the gluing process no large surpercoiling 
is involved so that 
 the angles ${\chi}_{in}$ and  $ {\psi}_{in} $ do not exceed $ 2\, \pi$. 
 As we will see
  the relevant  scaling variable for the partition function  is:
\be
   \eta= {\chi(L) A \over L}
 \ee
 In the 
experiments  to  be analyzed in the present paper, $ A/L  \simeq {1\over 300} $ 
 with $\eta$ of the order of unity,
so that ${\chi}_{in}$ can be safely ignored in comparison to $ 2\pi\,n $. In the 
following
we shall drop the $ L $ dependance in $ \chi(L) $ and write $  \chi=  2\pi\,n= 
\phi(L)+\psi(L) $.
The above equation does not imply that   $  \chi $  is a 
discrete variable: 
it is just for practical reasons that the experimentalists  perform measurements
 with  an integer number of turns.  On the other hand, if we were dealing
with a closed   DNA  chains instead of an open one, the supercoiling 
angle    $ \chi(L) $, which is here  an arbitrary real,  
would get replaced   by $ 2\pi\ L_k $   where the integer $ L_k  $ is
 the   topological  linking  number.

 \centerline{\it 1. A local writhe formula } 
We  make now an assumption which may look,  at first sight, somewhat
trivial  but   has, as we shall see, far reaching consequences:

{ \it  The Euler angles  $ \phi(s) $ and $ \psi(s) $
are regular enough functions of $s$ to allow the replacement  of $\chi $    
 by a line integral  involving the   sum  of  Euler  angles derivatives,
taken   along the trajectory  $ \Gamma $ of the tip  of 
the tangent vector  $\vt(s)$.}
\be
 \chi=  2\pi\,n= \phi(L)+\psi(L)=\int_0^L ds \ \left( \dot{\psi} +\dot{\phi}\right)
\label{chif}
\ee
It  is then convenient to introduce 
the total  twist  along  the chain $ T_w $ which  appears as
 a Gaussian variable in the partition function.
\be
T_w=  \int_{0}^{L} ds=   \int_{0}^{L} ds {\Omega}_{3}
 = \int_{0}^{L} ds  \left( \dot{\psi} +\dot{\phi} \, \cos\theta\right) 
\label{tw}
\ee
where ${\Omega}_{3}(s) $ is the projection of the instantaneous rotation vector $ \Om $
onto the unit tangent vector  $\vt(s)$,
  given in terms  of Euler angles in  Appendix A.
 We now {\it{define} } a `local writhe' contribution $ \chi_W $
by substracting the total twist  $ T_w $ from the supercoiling angle  $\chi $  
\be
 \chi_W=\chi-T_w= \int_{0}^{L} ds \ \dot{\phi} (1-\cos\theta) 
 \label{wr}
 \ee
The above decomposition  follows from our analysis of
 the empirical supercoiling angle 
and a   mere line integral manipulation. This is reminiscent of the 
decomposition of the linking number into twist and writhe for closed chains
\cite{geometry,boles}.

Formula (\ref{wr}) has been first obtained by Fain { \it et al.}
 \cite{fain} by adapting  
to the case of open strings a formula first derived by Fuller \cite{fullerform}
for  closed strings. We have decided to include an explicit derivation here
in order to be able to describe some subtleties in its use, which will
require some regularization procedure.
 Since $\vt(s)$ is a unitary vector, the curve 
  $ \Gamma $ lies upon the  sphere  $ S_2 $ and it  is parametrized  by  the spherical 
coordinates $ ( \theta(s), \phi(s) ) $.  This representation  is  well known 
to be  singular at the two poles
$ \theta=0,\theta=\pi $
 ( the choice of the $ z $ axis is 
not arbritrary but dictated by the experimental conditions: it coincides with
the rotation axis of the magnetic bead and gives also
the stretching force direction).
With the usual convention, $\theta$ is required
 to lie within  the interval $ [0 ,\pi] $. This condition  plays 
 an essential role in the quantum-like  treatment of the WLC and RLC continuous models,
to be discussed later on. When the trajectory  $ \Gamma $  passes through the poles
the restriction imposed upon $ \theta $ implies that the function $ \phi(s) $ 
suffers form a discontinuity of $ \pm \pi $ and this invalidates
 our derivation of formula  (\ref{wr}). To get around this difficulty,
we have to pierce  the sphere  $ S_2 $  at the 
two poles. The two holes are defined by 
two horizontal circles  with an arbitrary small  but finite  radius $ \epsilon $
 with their centers  lying on the $z$ axis. 
 Note that $ S_2$ has now the topology of a cylinder so that the
piercing of the sphere  $ S_2 $  provides the necessary  topological 
 discrimination between the two distinct
physical states 
$ ( \theta(s), \phi(s),\psi(s) ) $ and $( \theta(s), \phi(s)+2\pi n, \psi(s) ) $.
A careful evaluation of formula  (\ref{wr}), involving continuously
 deformed $ \Gamma $  trajectories drawn on the  pierced   sphere   $ S_2 $  
in order to avoid the poles, gives   results   in agreement with those given by the 
non local closed loop writhe formula \cite{geometry},
 while  a naive application of (\ref{wr}) would lead astray. 
\centerline{\it 2. Solenoid and plectoneme supercoil configurations}
To illustrate this point, we  shall  evaluate $\chi_W $  for  right handed
solenoid and plectoneme supercoil configurations, having arbitrary orientations.
 Let us begin with the simple case  where the solenoid superhelix 
 is winding around the  $ z  $ axis.
The tangent vector tip trajectory $ {\Gamma}_{sol}$ consists of $n$ turns along 
the horizontal circle cut upon the  unit sphere   $ S_2 $ by the horizontal plane
$ z= \cos {\theta}_0 $   ( $ \tan {\theta}_0 = \frac{ 2 \pi R}{ P} $  where 
$ R $ and $P$ are respectively the radius and the pitch of the superhelix.)
The writhe supercoiling angle $\chi_W ({\Gamma}_{sol} )$ is easily seen  to be 
given by $ 2 \pi\, n\,( 1- \cos {\theta}_0 )$. 

The plectoneme supercoil configuration
is composed of two interwound superhelices  connected by a handle. The helices have
 opposite axis but the same helicity. The plectoneme  $\vt$ trajectory   $ {\Gamma}_{plec} $
is   written as the union of three trajectories: 
$ {\Gamma}_{plec} = {\Gamma}_{sol}^1 \cup {\Gamma}_{han} \cup {\Gamma}_{sol}^{2}$.
${\Gamma}_{sol}^1$ is just the  right handed solenoid considered above.
$ {\Gamma}_{han} $ is associated with the handle connecting the two
superhelices and its contribution to $\chi_W $  can be neglected 
in the large  $ n $ limit. $ {\Gamma}_{sol}^{2} $ is the  tangent vector 
trajectory generated by the superhelix  winding down the  $ z  $ axis; it is given by 
performing upon the $ {\Gamma}_{sol}^1$ the transformation
defined  by the change of spherical coordinates :
$ {\phi}^{\prime}=- \phi \;, \: {\theta}^{\prime}= \pi - \theta $.
One readily gets $\chi_W ({\Gamma}_{plec} )$ by writing :
$\chi_W ({\Gamma}_{plec} ) = \chi_W ({\Gamma}_{sol}^1 )+\chi_W ({\Gamma}_{sol}^{2} )=
 - 4\pi\, n\,\cos {\theta}_0$.
 
We   proceed next  to the deformation 
of $ \vt $ trajectories $\chi_W ({\Gamma}_{sol} )$ and  $ {\Gamma}_{plec} $ 
by  applying  a rotation 
 about the $ y  $ axis of an  angle $\alpha$ which is going to vary 
continuously from 0 to $ \pi $.  We will find  that a proper use 
  of formula (\ref{wr}) leads to  rotation invariant  results in the limit of small
but finite angular cutoff.
As a first step, let us show that  $\chi_W (\Gamma )$ 
can be written  as the circulation  along $\Gamma$ of a magnetic monopole  vector potential.
This technical 
detour will also be  used in the  forthcoming derivation of 
the RLC model Hamiltonian. 

\centerline{\it 3. The local writhe  $\chi_W $ as  a
magnetic monopole  potential vector line integral }
 Our aim is to prove that the  writhe   supercoiled angle $ \chi_W$ can be written
 as the line integral $ \int d\phi \, A_{\phi}=  \oint  {\vA}_{m}(\vr) \,d\vr  $
 where $ A_{\phi} =  \(( 1- \cos\theta \))$  
 will be identified as the $\phi $ spherical component  
of the potential vector   ${\vA}_{m}(\vr)$ of
a magnetic monopole of charge unity. Following the well known Dirac procedure,
   we write  ${\vA}_{m}(\vr)$  as  
the potential vector of a thin solenoid of arbitrary  long length,
the so called  Dirac String $\cal{L}$, lying  along the negative half $z$-axis :
\be
{\vA}_{m}(\vr)=  \int_{ \cal L } d \vl \wedge {\grad}_r \frac{1}{| \vr-\vl|}=
 \frac{ \vz \wedge \vr}{r (r+\vz\cdot\vr)}
\label{monop1}
\ee
where $ \vl = u  \vz $ with $ -\infty < \, u \leq 0$.
 Using the above expression of  ${\vA}_{m}(\vr)$, one derives   the
two  basic equations which allow   the writing of  $ \chi_W $   as the circulation of the 
magnetic monopole  potential vector, valid  when $ r + \vz\cdot\vr > 0 $ : 
  \bea
 {\vA}_{m}(\vr)\cdot d\vr &= &  \(( 1- \cos\theta \)) d\phi   
\label{monop2} 
\\
 {\vB}_{m}(\vr) &= & \grad \wedge {\vA}_{m}(\vr)= \frac{\vr}{r^3}
\label{monop3}
\eea
where $ {\vB}_{m}(\vr) $ is  indeed the magnetic field produced by the magnetic 
charge unit.

Let us now define $ S_{+}(\Gamma )$ ( resp  $ S_{-}(\Gamma )$ )
 as the part of the sphere  $S_2$  bounded 
by the closed circuit $\Gamma$  which is run anticlockwise ( resp clockwise ) around the surface
 normal. ( Note that  $ S_{+}(\Gamma ) \cup S_{-}(\Gamma )= S_2$ ).
Let  us call  $ {\Sigma}_{\epsilon} $ the part of  $S_2$ defined by the south pole hole
 and   assume that  $ S_{+}(\Gamma ) \cap   {\Sigma}_{\epsilon} =\emptyset $  ( as a consequence:
$ S_{-}(\Gamma ) \cap  {\Sigma}_{\epsilon} = {\Sigma}_{\epsilon} $). Applying
the Stokes theorem  to formula ( \ref{monop3} ),
we can write:
\be 
\chi_W (\Gamma)= {\oint}_ {\Gamma } {\vA}_{m}(\vr)\cdot d\vr= 
 {\int \int}_{S_{+}(\Gamma )} d \vS \cdot  {\vB}_{m}(\vr)= {\cal A}( S_{+}(\Gamma ))
\ee
where ${\cal A}( S_{+}(\Gamma ))$ is the area of the spherical cap $ S_{+}(\Gamma )$. 
The closed circuit $\Gamma$ is assumed  here to be run only once; if it is run 
$n$ times, as in solenoid and plectoneme configurations, $\cal{A}( S_{+}(\Gamma ))$
should then be multiplied by $n$ and the formulas obtained before are immediately recovered.
 A similar formula, valid {\it mod} $2 \pi $,
has been derived previously  by Fuller \cite{fullerform}, in the case of
  closed molecular chain.
\centerline{ \it 4. On the rotation invariance  of the writhe supercoiling angle $\chi_W$}
 We are going  now  to follow the variation of $ \chi_W (\Gamma) $
when the  initial tangent vector  $\vt$  trajectory   ${\Gamma}_0$ is
 moved continuously upon the sphere   by  applying  a  rotation 
 about the $ y  $ axis by an  angle $\alpha$, varying continuously from 0 to $ \pi $:
$  {\Gamma}_0  \Longrightarrow  \; \Gamma(\alpha)= R(\vy ,\alpha) {\Gamma}_0  $.
We shall choose  $ {\Gamma}_0 = {\Gamma}_{sol} $ but the following analysis can
be easily extended  to plectonemes or more general configurations 
involving  closed $ \vt $  trajectories.
 
 The crossing  of the north pole is harmless because $ 1-\cos\theta $ 
vanishes at $ \theta= 0 $ and  everything goes smoothly until $ \Gamma(\alpha)$
 meets the south pole hole ( $ \theta= \pi$ ) when $\alpha$ is approaching $ \pi -{\theta}_0$.
When     $ \alpha < \pi-{\theta}_0 $ the south pole stays outside $ S_{+}(\Gamma( \alpha ))$ and
${\chi}_W (\Gamma)$ is given by $n$ times the spherical cap area: 
 ${\cal A}( S_{+}(\Gamma(\alpha ))=
 {\cal A}( S_{+}(  {\Gamma}_0  )) = 2\pi ( 1- \cos {\theta_0} )$.
In other words   $  \chi_W (\Gamma(\alpha) ) = \chi_W (  {\Gamma}_0) $ if
  $ \alpha < \pi-{\theta}_0 $.
 
  In order  to avoid the  hole $ \Gamma(\alpha) $ 
 must be  continuously  deformed  for $ \alpha > \pi-{\theta}_0 $
into  the circuit ${\Gamma}^{\prime}(\alpha)= \Gamma(\alpha) \cup  {\Gamma}_{\epsilon} $ 
where  the path $ {\Gamma}_{\epsilon} $
consists  in  $n$ turns  run anticlockwise  around  the south pole hole of radius $\epsilon$. 
(We ignore the two ways narrow lane  connecting the two loops which gives a vanishing
contribution.)   It is clear now that $ S_{+}(\Gamma(\alpha ) $  contains 
the south pole,  so we must  apply the Stokes theorem to 
the clockwise spherical bowl  $  S_{-}(\Gamma(\alpha ))$:
$$
{\chi}_W (\Gamma(\alpha))= 
 -n  {\int \int}_{S_{-}(\Gamma (\alpha  ))} d \vS \cdot
  {\vB}_{m}(\vr)= -n {\cal A}( S_{-}(\Gamma(\alpha ))
= n ({\cal A}( S_{+}(\Gamma(\alpha ))- 4 \pi)
$$
The contribution of the path  ${\Gamma}_{\epsilon}$ is readily  found 
to be $  4 \,\pi \, n (1+ O({\epsilon}^2 ) )$
so it just cancels the $ - 4 \,\pi \, n $ term  in  the above result.
  As a consequence,  
 the writhe supercoiling angle associated with the south pole avoiding $\vt$ trajectory 
${\chi}_W ({\Gamma}^{\prime}(\alpha))$
coincides when $ \pi-{\theta}_0 <\alpha < \pi $ with  the $ \alpha=0 $ initial result:
${\chi}_W ({\Gamma}^{\prime}(\alpha))=  \chi_W (  {\Gamma}_0)= 2\pi n ( 1- \cos {\theta_0} )  $
up to corrections $ O({\epsilon}^2 ) $. 

{\it  In conclusion, we have shown that  a proper use  of formula (\ref{wr}), in  the simple case
of  solenoidal configurations,  leads to rotation invariant $\chi_W$
in the limit  of small $\epsilon$. } 
 
This result  can be easily extended to 
 the plectonemic  configurations and even to more general ones. 
The crucial cancellation is  taking place in the vicinity of the south pole crossing
and thus  does not depend on the detailed shape  
 closed trajectory  ${\Gamma}_0$ in the $\vt$ space.
If one adds the minimal  extra  string section 
to the solenoid or  plectoneme, necessary  to get a closed loop  in  
the  physical space, our findings are in  agreement, in the limit of large $n$,
with  those  obtained  from the  non-local closed loop
 writhe formula \cite{geometry}, which is  explicitly rotation invariant from the start.
   Formula (\ref{wr}) is thus a correct description of the problem provided 
the trajectory is obtained
from the straight line $ (\theta(s)=0 ,\phi(s)=0) $ by a continuous 
transformation on the unit sphere
pierced at $\theta=\pi$.
 A similar condition was obtained by Fuller \cite{fullerform} in his local Writhe formula
for closed strings. An alternative derivation of our result, more similar to the 
approach of Fain \cite{fain}, could be to close the open string by some 
straight line closing at infinity and rely on Fuller's formula,
but this procedure raises subtle questions about the contribution of added pieces
which made us prefer the present explicit derivation.

\begin{center}  
\subsection{ The Rod Like Chain Hamiltonian}
 \end{center}

The partition function for a fixed value of
$ \chi $  is given by the  path integral in the space of Euler angles :
\be
Z(\chi,F) =\int { \cal D }(\theta,\phi,\psi) \,
 \delta \((\chi-\int_0^L ds (\dot{\phi}+\dot{\psi})\)) \exp- {E_{RLC}\over 
k_{B}T}
\label{pathint}
\ee
 The functional space $ \theta(s),\phi(s),\psi(s) $  should be specified in accordance 
with  the previous considerations, in particular the tangent vector $ \vt $ paths
 must bypass the holes  on the pierced sphere   $ S_2 $. This can be achieved 
by introducing explicitly  in $ E_{RLC} $  a short range repulsive potential near $ \theta= \pi $.

Two   extra constraints, non local in 
the tangent vector $\vt$ space should have been, in principle, implemented
in the above partition function.

The first one concerns
self avoidance effects induced  by  the screened Coulomb repulsion
between different molecular sections. Even the simple
version of this constraint, used in supercoiled  plasmids 
Monte Carlo simulations\cite{volog}, is exceedingly  difficult 
 to implement in the Quantum Mechanics inspired formalism
of the present paper. Here Coulomb 
 induced self avoidance wail be ignored  as  all   quantitative analytic
 approaches did so far,
including the celebrated   WLC computation.
It will appear that these effects 
  play  a limited role in the low 
supercoiling regime ( $\vert\sigma \vert \leq 0.02 $)  to be analyzed in the present
paper,
a regime and experimental conditions 
 where the WLC model is   also working beautifully.

The second  non-local constraint  not implemented in the present work
 concerns the exclusion of knotted 
chain configurations. In principle, knotting 
an open DNA chain  is not forbidden, but such a transition is
expected to be  inhibited on
the time scale of the actual experiment since the contour length
of the  molecule $L$ is only 1.7 times the 
circumference of the  bead glued at the free end of the molecule.

Using a Fourier representation of the Dirac $  \delta  $ function we can write 
the partition function  as a Fourier integral:
$  Z(\chi,F)=\int dk \exp(-ik \, \chi )  \tilde Z(k) $ where 
the Fourier transform $   \tilde Z(k) $ is given by the functional integrals:
\begin{eqnarray}\nn
 \tilde Z(k) &  = & \int { \cal D }(\theta,\phi) \,\exp  \((
i\,k\, \chi_W- {{\cal E}_{bend}+{\cal E}_{stretch}
\over k_{B}T} \)) {\tilde Z}_{T}(k)  \label{Ztild} \\ 
{\tilde Z}_{T}(k) & = &  \int { \cal D }(\psi) \,\exp \(( i\, k\, T_{w} 
-{{\cal E}_{twist} \over k_B T}\))
\label{ZtildT}
 \end{eqnarray} 
where  the above factorization follows from  the identity:
$\int_0^L ds (\dot{\phi}+\dot{\psi})= T_w +\chi_W$, using the elastic energy densities
defined in (\ref{EL}).
The  Gaussian path integral upon $ \psi $ is readily performed and gives the 
Fourier transform of the twist partition function of equation (\ref{ZT})
$  {\tilde Z}_T(k) = \exp( -{k^2L\over 2 C})$, 
up to a trivial constant. The partition function Fourier
 transform ${\tilde Z}(k)$ can then be written as  the product: 
$  {\tilde Z}(k)= {\tilde Z}_T(k) {\tilde Z}_W(k) $
where $ {\tilde Z}_W(k) $, which is interpreted
 as the writhe partition function Fourier transform,
 is given by a path integral upon $\theta $ and $ \phi $, with the effective energy:
\be 
{\tilde E}_W (k) =  {E_{WLC}\over k_{B}T} +i\,k\, \chi_W
 \label{Etild}
 \ee 
Using the results of the previous section ones sees immediately that 
$ i\,k\, \chi_W $ can be written as $i$ times the line integral 
$ \int{\vA}_{m}(\vr) \,d\vr  $
where  ${\vA}_{m}(\vr)$ is  the vector potential produced by  the magnetic monopole, having now
a charge $k$. 
 If one performs, as in Section II.B, an
analytic continuation of  the integral over $s$  in $ {\tilde E}_W (k) $
 towards the imaginary axis   the $ i $ factor disappears
 in the  writhe line integral.
  One arrives to  the action integral of a unit charge  particle
 moving  on the unit sphere under the joint action of an electric field $ f$
and  a magnetic monopole of charge $k$.
In order to get the corresponding Hamiltonian ${\hat H}_{RLC}(k)$ we
apply a standard  Quantum Mechanics rule  used to 
implement the switching on of a magnetic field:
we replace in the kinetic terms $ \frac{1}{2 A} {\vp}^2 $,  the particle momentum $\vp $ 
by $\vp-e{\vA}_{m}(\vr) $ (here $e$=1).
The term linear in ${\vA}_{m}(\vr)$ disappears  by averaging over the final angle $ \phi = \phi(L)$
  and we are left  with the diamagnetic term:
\be
  \frac{1}{2 A}  {\vA}_{m}^2(\vr)=\frac{k^2}{2A}{1-\cos\theta \over 1+\cos\theta} 
\label{diam}
 \ee
Adding (\ref{diam}) to the    WLC Hamiltonian, one arrives  to the dimensionless
 Hamiltonian ${\hat H}_{RLC}(k)$ associated with the partition function ${\tilde Z}_W(k) $:
 \be
{\hat H}_{RLC}(k) =
-\frac{1}{ 2\,\sin\theta }\frac{\partial}{\partial\,\theta}\sin\theta \,
  \frac{\partial}{\partial \theta }-
\alpha \cos\theta+{k^2 \over 2} {1-\cos\theta \over 1+\cos\theta}
\label{HRLC}
 \ee
 where we have taken, as the length unit, the persistence length $ A $   and introduced 
the dimensionless force parameter $\alpha=  {F A  \over  k_{B} T}$. 

{ \it Let us stress that the above Hamiltonian is a straightforward
consequence of the expression for the elastic energy ( Eq \ref{EL0},\ref{EL}) and our
expression (\ref{chif}) giving  the supercoiling 
constraint $\chi$ in terms of  Euler angles derivatives. }

Moreover, as it is shown in Appendix B, ${\hat H}_{RLC}(k)$  can be derived  directly,
{\it without having to introduce the local writhe given by  formula (\ref{wr})}. 
The method involves  the quantization  of the  symmetric top Hamiltonian as
 a differential operator   acting upon
  non-periodic  functions of the  Euler angles $ \phi $ and $\psi$.
This is required to
 describe  the thermal  fluctuations of an elastic rod ( see  section II.B) and $k$
 appears  as the continuous angular momentum along the top axis.

 Introducing  the eigenstates and the eigenvalues of ${\hat H}_{RLC}$,
$ {\hat H}_{RLC}{\Psi}_{n}(k,\theta)= {\epsilon}_n(\alpha,k^2) {\Psi}_{n}(k,\theta)$,
 the Fourier transformed partition function $\tilde Z$ can be written as the
sum:
\be
\tilde Z=\sum_n { \Psi}_{n}(k,\theta_0){\Psi}_{n}(k,\theta_L)
\exp\((- \frac{L}{A} \(({\epsilon}_n(\alpha,k^2)+ {k^2 A \over 2 C }\))\))
\ee
(Note that ${\hat H}_{RLC}$ being a real operator the 
eigenfunctions can always be  taken as real.)

In the large $L$ limit, the sum over the eigenstates is dominated by the one
 with lowest energy $\eps_0(\alpha,k^2)$, if $ L/A \gg \Delta\eps $
where $\Delta\eps$ is the energy
gap between the ground state and the nearest excited state of  ${\hat H}_{RLC}$.
  This gives the approximate expression for the partition function
$ Z ( \chi,F) $ :
\be
Z  \simeq \int dk \  \exp\((- \frac{L}{A} \((  {\epsilon}_0(\alpha,k^2) +
{k^2  A \over 2 C} \))-i\, k\,\chi  \)) 
\label{Zchi} 
\ee
To get the above equation, we have also  neglected in $ \tilde Z(k)$
the prefactor involving the ground state wave function:
 ${ \Psi}_0(k,\theta_0){\Psi}_0(k,\theta_L)$. It leads to
finite size  corrections of order  $ A/L $, which are in general dominant 
with respect to those associated with the excited states, which
scale as $ \exp( - \Delta\eps L/A )$. 

\subsection{ The pathology of  the RLC model in the continuous limit.} 
In order to comply with the prescription given  for the path integral
(\ref{pathint}) we should have added  to ${\hat H}_{RLC}(k)$ a strong repulsive 
potential acting in the interval: $ \pi -\epsilon < \theta <\pi $, in order to avoid the 
 $ \theta =\pi $ singular point. 
We shall refer to the $\eps \to 0$ limit as the 'continuous limit',
 and show that in
that limit the model is pathological.

 Let us derive from the  ground state energy 
$\eps_0(\alpha, k^2)$
of the Hamiltonian ${\hat H}_{RLC}$,  some observable properties of a long RLC.
In the following the variable $ z =  k^2 $ is assumed to take  positive real values.

 If instead of constraining $\chi$
 one  measures its thermal fluctuations,
their probability distribution is just $P(\chi) \propto Z$. Let us 
consider  the second moment as we did in  subsection II.B for the case 
of a non-flexible rod.
\bea
   <\, {\chi}^2 \, > & =-& \lim_{k^2\to 0}{1\over\tilde Z (k)}
 { \partial^2 \tilde Z (k)\over\partial\, k^2 } \\
                     & = &   { L\over C}+
  { 2\, L\over A} \lim_{k^2\to 0}{ \partial \epsilon_0(\alpha,k^2) \over
\partial\, k^2 }
\label{chi2}
\eea
We recognize in the first term of equation  (\ref{chi2})  the 
contribution  to
$<\chi^2>$  from the twist fluctuations, ${ L\over C}$, obtained previously.
The second piece of (\ref{chi2}) gives the contribution 
from $\langle \chi_W^2 \rangle$, but it turns out to be   divergent. 
Evaluating
$\epsilon_0(\alpha,k^2)$ at
small $k^2$ from standard perturbation theory, we find
$ \langle{\chi_W}^2 \rangle
    =  (L/A) \langle \Phi_0\vert(1-\cos\theta)/( 1+\cos\theta)\vert 
\Phi_0\rangle
$
where $\Phi_0(\theta)$ is the groundstate eigenfunction  
   of the WLC Hamiltonian (which is ${\hat H}_{RLC} $ at $ k=0 $). As
 $\Phi_0(\theta=\pi)\ne 0$ (for any finite force), we get clearly  a logarithmically 
divergent result.
To study analytically this peculiar theory two methods have been followed,
associated with  different limits.

 {\it i)The weak force limit}: 

 When $ \alpha =0 $   the eigenvalue problem 
for $ H_{RLC}$ 
can be solved  exactly. The eigenfunctions are given
 in terms of Jacobi polynomials : 
$${\Psi}_n(k,\theta)= (1+\cos\theta)^{k} P_n^{(0,2 k)}(\cos\theta)$$
and the  associated  eigenvalues are:
 $$ {\eps}_0( \alpha=0,k^2)= \(( (2 n+1) k +n^2+n \))/2 $$ 
 In the presence of a small force ($\alpha \ll 1$), the 
 ground state energy shift is easily calculated to second order in $ \alpha $: 
$$ \delta_2 {\eps}_0(\alpha,k^2)= -{\alpha k\over 1+k} -
 {{\alpha}^2 \over (1 +k)^3 (3 + 2 k )} $$
 Note that $ {\eps}_0(\alpha,k^2 ) $    is not an analytic function 
of $ z=k^2 $. 

{\it ii)The small $k $  limit}:
In order to build a perturbative expansion near $k  =0 $ it is 
convenient 
 to factorize in the wave function  the singular term near $\theta=\pi $ 
by writing :
${\Psi}_0(k,\theta)= (1+\cos\theta)^{k} {\chi}_0(k,\theta)  $. 
The wave function  $\chi_0(k,\theta)$ obeys a new wave equation 
which is amenable to a well defined perturbation expansion in $ k $. 
It 
will turn out that  the  values of $ k $ relevant for 
the evaluation of Fourier transform  are of the order of $ A/L$, so that a first
order perturbation is sufficient (at least when  $ \alpha $ is small enough in 
order
 to guarantee that $ \Phi_0(\pi)^2 $ stays of the order of unity). 
We have found that  the ground state energy is  given  to first 
order in 
 $ k$ by:
$ \eps_0(\alpha,k^2) = \eps_0(\alpha,0)+ k \Phi_0(\pi)^2+ O(k^2) $.
It can be checked that the perturbation  result of method {\it i)}  
   is   consistent with the above equation. 
 The Fourier integral  can be  readily performed,  leading to a   Cauchy type  
 probability distribution,
  which is best expressed in
 terms of the reduced variable 
$\eta=\chi A/L$:
 $ P(\eta ) \simeq {1\over \pi}\,{\Phi}_0(\pi)^2 /( {\Phi}_0(\pi)^4 + \eta^2 )$
 and thus leading to a  diverging second moment.This last point can also be seen from the fact 
that ${ \partial \epsilon_0(\alpha,k^2) \over\partial\, k^2 } \propto {1\over k} $
in the small $k$ limit.
The relative extension
of the  chain in the direction of the force is given by
${\langle z \rangle / L} = (A / L) {\partial \ln Z \over \partial \alpha}$.
The partition function in the present case is easily found to be : 
$ Z( \eta,\alpha) \propto \exp\(( -L/A\, \eps_0(\alpha,0) \)) \,  P(\eta ) $. 
It leads to:  $ {\langle z \rangle / L} =
 -{\partial \eps_0 \over \partial \alpha} (\alpha,0) +O(A/L) $, {\it i.e.} the 
same result as the WLC model. 

 { \it The present  computations 
show that the continuous RLC model leads to predictions 
which are both pathogical ( infinite second moment of the  writhe spontaneous fluctuations )  and in striking contradiction with experiment (  absence of variation of 
the average extension with respect to the supercoiling angle  $\chi$ at moderate
forces ). All these peculiar features have been confirmed with explicit computations
using a regularized $ H_{RLC}$ hamiltonian. We have indeed found that 
$< \chi_{W}^2 >$ scales like 
$ {L \over A} \log{1 \over \epsilon} $
 and  that the 
extension versus supercoiling  curves flatten up in the small $\epsilon$ limit.}

As a last remark we would like to note that if the variable $z$ is continued 
analytically towards negative value $ z \rightarrow -{\kappa}^2$ 
the groundstate  energy  develops an imaginary part.This a clear
indication that  ${\hat H}_{RLC}( i\kappa ) $ has no stable 
ground state.

 
\begin{center}
\section{  Discretization as a Regularization Method  for the Rod Like Chain  Model }
\label{discrete_theory}
\end{center}

The introduction  by hand of  an angular cutoff near $ \theta =\pi$ 
 appears as an {\it ad hoc } procedure. We are going to show in the present section 
that a discretization of the chain involving an elementary link of length  
$b$ generates an angular cutoff: $ \sin^2\theta  \geq {b\over A} $.
 We have  seen in section II.A that the RLC 
model with its cylindrical symmetric rigidities  tensor is realistic only in presence of a 
finite resolution $ \Delta \,l $ in the length measurement. The existence of a length 
cut-off $ b \sim  \Delta \,l $  appears not only  natural but necessary
in the present context. The non trivial
fact is that the  "mesoscopic " elastic properties,  taking place on the length scale 
of the  whole supercoiled   molecule (about  ten microns), are
sensitive to  the existence or not of a cutoff in the range of few nanometers.
We should also stress that length cutoff effects are found to
 be reduced to the level of few percents when supercoiling is absent. 
  Therefore in that case
 the continuous version of the  WLC model remains  a remarkably good description 
of the  DNA force extension measurements. 

Moroz and Nelson \cite{mornel1}  have  devised a computation procedure where 
no angular cutoff is introduced from the start. They consider the
situation where the torque acting upon the molecular end is kept 
fixed so that  the relevant Hamiltonian is just ${\hat H}_{RLC}( i\kappa ) $.
The pathology discussed above is expressed in the fact 
that this Hamiltonian   has no stable  ground state.  In order to deal with such a 
situation, they  use a perturbation method near  the high force
limit where the relative elongation is close to unity. More
precisely, they construct a perturbation expansion involving  negative power of 
$ K= \sqrt{\alpha - {1\over 4} {\kappa}^2 }$.  The  lowest order
Hamiltonian  potential is given, up to  a constant, by 
 the  small angle approximation:
 $ {1\over 2} K^2 {\theta }^2={1\over 2} ( \alpha - {1\over 4} {\kappa}^2 )  {\theta }^2 $.
Since    ${\hat H}_{RLC}( i\kappa ) $ is not a self adjoint operator, 
the series
diverges and it is expected  to be, at best, an asymptotic series. 
The regime of validity of the prediction is difficult to assess in this approach 
since one is basically approximating a divergent ground state energy from
the first terms of the perturbation series.
As long  as one sticks to a finite order, the $ \theta =  \pi $ singularity does not show up.
    In  order to make contact with experiment,
Moroz and Nelson  have been forced to restrict their analysis to
the domain  of force and supercoiling
where  $ K > K_c=3 $. 
In our approach, we have instead regularized the model explicitely,
as we shall discuss. This allows to get a better control of the
theory and it turns out that our results hold on a much wider range of force-supercoiling.
It can even be compared
in the $ F=0 $ limit  to supercoiled plasmids experimental  data and Monte Carlo simulations.
( See  Section \ref{tw_wr_pl} ).

 \subsection{  Construction of the transfer matrix associated with  the 
discretized 
 RLC model }
The continuous elastic RLC is  transformed into  a chain composed of $ N $ 
elementary
 links of length b. The  continuous arc length $ s$ is replaced by a discrete 
set:
$ s_n = n \,b $ with $ 0 \leq n \leq N $  and $ L= N b $. We have  now to 
specify the   
discretization  rules. For the   bending linear energy density 
${\cal E}_{bend}$, there is   a  rather natural prescription:
 it is to replace $ \vert  {d \vt(s) \over ds} \vert ^2 $  by 
$ {1 \over b^2} { \vert {\vt}_{n} - {\vt}_{n-1} \vert}^2  $.
 
 We are led in this way  to {\it the first discretization rule}: 
 \be
 {\cal E}_{bend}  \Longrightarrow  { A \over b^2} \lbrack 1-\cos( {\theta}_n - 
{\theta}_{n-1} )+
( 1-\cos( {\phi}_{n} - {\phi}_{n-1} )\,\sin {\theta}_{n} \, \sin {\theta}_{n-1} 
\rbrack
\label{rule1}
\ee 
The above discretized linear density is endowed with remarkable  feature: 
it is  $ 2\pi $-periodic with respect to the angular finite
 difference $ {\phi}_n - {\phi}_{n-1} $. This property plays an 
 essential role  in the derivation  of the transfer matrix $ T({\theta 
}_n,{\theta }_{n-1})$ which
allows a direct computation of the  partition function ${\tilde Z}(k)$, averaged 
upon the 
azimuthal angle $ \phi(L)$.

For the term generated by the supercoiling constraint: $ ik (1-\cos\theta(s) 
)\dot{\phi} $, we
 shall impose the same periodicity  condition, together with a symmetry under 
the exchange 
$ {\theta }_{n+1} \Leftrightarrow  {\theta}_n $, in order to insure the 
hermiticity of the
 transfer matrix.
 
We arrive in this way to {\it the second discretization rule}:
\be
\(( 1- \cos \theta \)) \,  \dot{\phi}  \Longrightarrow  { 1 \over b} \sin ( 
{\phi}_n - {\phi}_{n-1} ) \,
\(( 1 - { \cos {\theta}_n +\cos {\theta}_{n-1} \over 2} \))
\label{rule2}
\ee
The discretized version ${\tilde Z}_{N}(k,{\theta}_{N},{\phi}_{N})$ of
the writhe partition function 
can be written as 
the following $2 N$ dimension integral:
$$
{\tilde Z}_{N}(k,{\theta}_{N},{\phi}_{N} )=\int\,\prod_{n=1}^{N} d{\phi}_{n-1}\,
 d(\cos{\theta}_{n-1})\,
\exp -b\,\(( \Delta {\cal E}( k,{\theta}_n,{\theta}_{n-1},{\phi}_n-{\phi}_{n-1})\))
$$ 
where $\Delta {\cal E}( k,{\theta}_n,{\theta}_{n-1},{\phi}_n-{\phi}_{n-1}) $ is the  
discretized
linear energy density  obtained from
(\ref{Etild})
 by the replacement rules (\ref{rule1}) and 
(\ref{rule2}). 
It is convenient to introduce the partition function $ {\tilde 
Z}_{n}(k,{\theta}_{n},{\phi}_{n} )$
corresponding to the intermediate value $ s_n= b n $. 
One gets then the recurrence relation:
\begin{eqnarray}
{\tilde Z}_{n}(k,{\theta}_{n},{\phi}_{n})
 &=& \int  d{\phi}_{n-1}\,d\((\cos {\theta}_{n-1}\)) \,
 \exp -b \,\(( \Delta {\cal E}( 
k,{\theta}_n,{\theta}_{n-1},{\phi}_n-{\phi}_{n-1})\)) \nonumber \\
       & &  {\tilde Z}_{n-1}(k,{\theta}_{n-1},{\phi}_{n-1})
\label{recur1}
\end{eqnarray} 
Since we are ultimately interested in the  partition function ${\tilde Z}(k)$ 
averaged upon the 
azimuthal angle $ \phi(L)$, we shall define :
\be
z_n(k,{\theta}_n)= {1 \over 2 \pi M }\int_{ -\pi\, M}^{ \pi\, M} d{\phi}_{n} \,
                                        {\tilde 
Z}_{n}(k,{\theta}_{n},{\phi}_{n})
\label{znav}
\ee 
where  $ M $ is an integer which  can be arbitrarily large. We perform the  above 
integral upon
$ {\phi}_{n} $ in the two sides of equation (\ref{recur1}). In the right hand side we 
change the  integration variables $  {\phi}_{n} , {\phi}_{n-1} $ into the new set:
 $u_n=  {\phi}_{n} - {\phi}_{n-1},{\phi}_{n-1} $. Using the periodicity  of
the integrand with respect to $ u_n $,
we can  make the replacement:
 $$
{1 \over 2 \pi M } \int_{ -\pi\, M - {\phi}_{n-1} }^{ \pi\, M- {\phi}_{n-1} } 
du_n ...
  \Longrightarrow      {1 \over 2 \pi  } \int_{ -\pi }^{ \pi} du_n ...
$$ 
The integral upon $ u_n $ can then be performed exactly in terms of the  Bessel 
function
of second type $ I_0(z) $
and we arrive finally to an explicit recurrence relation involving the
$ \phi$-average partition function  $z_n(k,{\theta}_n)$ :

\begin{eqnarray}
  z_n(k,{\theta}_n) &=&  \int_0^{\pi}  \sin {\theta}_{n-1} d{\theta}_{n-1}\,
    T({\theta }_n,{\theta }_{n-1} )\,z_{n-1}(k,{\theta}_{n-1})\label{recur2}  \\
 T({\theta }_1,{\theta }_2,k^2) & =& \exp-\lbrace \frac{A}{b}\, \left( 1 
                                  - \cos ({\theta }_1-{\theta }_2 ) +
                                  \sin {\theta }_1\, \sin {\theta }_2  \right) 
\nonumber \\
 & &  +\frac{b\,\alpha }{2\,A} \left( \cos {\theta }_1 + \cos {\theta }_2 
\right)  \rbrace
 \,I_0 (f({\theta }_1,{\theta }_2,k^2))
\label{mtrans}
\end{eqnarray} 
  where the Bessel function argument  is given by :
\be 
f({\theta }_1,{\theta }_2,k^2) = \sqrt{ -k^2 \,{ \left( 1 -  \frac{\cos {\theta }_1 + 
                \cos {\theta }_2 } {2} \right) }^2  + 
     \frac{ A^2}{ b^2} {\left( \,\sin{\theta }_1\, \sin{\theta }_2 \right)}^2  } 
\label{besarg}
\ee  

  To get some basic elements of the transfer matrix formalism it
 is useful to rewrite the recursion relation  (\ref{recur2})
 within an operator formalism: \\
$ \vert z_{n} \rangle = \hat{ T }(\alpha,k^2)\vert z_{n-1} \rangle $
 where  $\hat{ T }(\alpha,k^2)$ is the operator 
associated with the transfer matrix $ T({\theta }_1,{\theta }_2,k^2)$. Let us 
define as 
 $ \vert t_i(\alpha,k^2) \rangle $ the eigenstate of $\hat{ T }(\alpha,k^2)$ 
associated with the 
eigenvalue $  t_i(\alpha,k^2) $. Performing  an expansion   upon the  above 
set of eigenstates, 
we can write   $ \vert z_N \rangle $ as follows:
$$
 \vert z_N \rangle  = {\hat{ T }(\alpha,k^2)}^N \,\vert z_{0} \rangle
= \sum_{i} { t_i(\alpha,k^2)}^N \, \vert t_i(\alpha,k^2) \rangle 
 \langle t_i(\alpha,k^2) \,\vert z_{0} \rangle
$$ 
Let us assume that $N$ is large enough so that the above
sum is dominated  by the contribution from the lowest eigenvalue.
 The partition function $ Z(\chi) $ can then be written as: 
$$ 
 Z(\chi)= Z  \simeq \int dk \  \exp\((- {k^2 \over 2 C} +i\, k\,\chi  \)) \, 
{t_0(\alpha,k^2)}^N 
$$
It is now convenient to introduce the { \it  new definition }:
\be 
  \eps_0(\alpha,k^2) = - {A \over b}\, \ln t_0(\alpha,k^2)
\label{epstr}
\ee 
Note that we had already defined    $  \eps_0(\alpha,k^2) $ as the lowest
eigenvalue  of $ H_{RLC} $; it is easily verified that the two definitions 
coincide in the limit $ b/A \ll 1$ {\it modulo } an irrelevant constant. 
With this new definition the equation (\ref{Zchi}) giving  $ Z(\chi) $ holds true 
for  the 
discretized RLC model, within the transfer matrix formalism.
\subsection{  The angular  cutoff induced by the discretization of the RLC model 
}
In this section we are going to discuss what turned out to be a crucial 
milestone in
our work: the understanding  that a discretization 
of the DNA  chain  could provide   the angular cutoff 
needed  to make sense to  the RLC model.
 The first indication came, in fact, from 
MonteCarlo simulations  which will  be discussed later on in section  \ref{Monte_Carlo}. 
We would like to   show here that, indeed, the angular cutoff comes out from   
the transfer matrix  solution of the discretized RLC model. 
Moreover our analysis will lead
us to the formulation  of a regularized version of the continuous  
RLC model.  
 
Our procedure  involves an  analytic computation of the 
partial derivative with respect to $ k^2 $ of 
the transfer operator ground state energy  $\eps_0(\alpha,k^2)$, 
as defined  by equation (\ref{epstr}) :
   $\partial_{k^2} {\epsilon}_0=  { \partial \epsilon_0(\alpha,k^2) \over
\partial\, k^2 }$. 
 This derivative is very important physically, for the 
following two reasons.
  First, we remind that   $ \lim_{k^2\to 0} \partial_{k^2} {\epsilon}_0$ 
gives  the second moment of the 
 spontaneous "writhe" fluctuations $ \langle {\chi_w}^2  \rangle $ 
 (see equation (\ref{chi2}) ), which was
found to be infinite within the  continuous RLC model.
Second,  for  imaginary finite  values of
$ k=i \kappa $, $ \partial_{k^2} {\epsilon}_0$   enters in an essential way   
in the  parametric representation  of the extension 
versus supercoiling  curves, to be derived  in the next sections.

The partial derivative $ \partial_{k^2} {\epsilon}_0 $ is expressed,
 using   Eq.(\ref{epstr}), as
\be 
 { \partial \epsilon_0(\alpha,k^2) \over \partial\, k^2  }  =
{-A \over b\,  t_0(\alpha,k^2) } \, { \partial t_0(\alpha,k^2)\over \partial\, k^2 }
=  {-A \over b\,t_0(\alpha,k^2) } 
\langle t_0 \vert {\partial\hat{ T }(\alpha,k^2)\over  \partial\, k^2 } \vert t_0 \rangle 
\label{depsk2}
\ee 
The partial derivative with respect to $ k^2$  of the
transfer matrix $ T({\theta }_1,{\theta }_2,k^2) $ is
 easily obtained  from equations (\ref{mtrans}) and (\ref{besarg}):
\begin{eqnarray}
 {\partial T({\theta }_1,{\theta }_2,k^2) \over  \partial\, k^2 } & = &
T({\theta }_1,{\theta }_2,k^2) {\cal W}({\theta }_1,{\theta }_2)  \nonumber \\ 
{\cal W }({\theta }_1,{\theta }_2) & = &
-{1\over2}  R( f({\theta }_1,{\theta }_2,k^2) ) 
 { { \left( 1 -  \frac{\cos {\theta }_1 + 
           \cos {\theta }_2 } {2} \right) }^2 \over f({\theta }_1,{\theta }_2,k^2)} 
 \end{eqnarray}
 We have defined  the small $x$ cutoff function 
\be
  R(x)= {I_1(x)\over  I_0( x)} 
\label{Rdef}
\ee  
  which  behaves like $x$  
when $ x \ll 1$  and goes rapidly to 1 when $ x >1 $.
The next  step is to investigate the variation 
of $  T({\theta }_1,{\theta }_2,k^2) $ as  a function 
of $ \Delta\theta = {\theta }_2 -{\theta}_1 $ for a fixed value of
  $  \theta =({\theta }_1+{\theta }_2)/2 $
when $ A/b \gg 1 $.  
 For the sake of simplicity  we limit ourselves to value of $ k^2 $
such that $ \vert   k^2 \vert {b^2\over A^2 }\ll 1 $ 
 so that $ f( \theta_1 ,\theta_2 ,k^2 ) $ can be
approximated  by $ {A \over b} \sin\theta_1  \sin\theta_2 $. 
The transfer matrix is  then given 
by the somewhat simplified expression:
\be
T({\theta }_1,{\theta }_2,k^2)  = 
 \exp-\lbrace \frac{A}{b}\, \left( 1 - \cos ({\theta }_1-{\theta }_2 )  \right) +
 \frac{b\,\alpha }{2\,A} \left( \cos {\theta }_1 + \cos {\theta }_2 \right) \rbrace
  B_0( \frac{A}{b} \sin {\theta }_1\, \sin {\theta }_2 ) 
\ee
where we have  introduced the auxiliary function:
$ B_0(x)= \exp(-x) I_0( x) $ which  is slowly varying with $x$: constant near the origin
it behaves like  $ 1/\sqrt{x} $ for large $x$.
This formula shows  that  $  T({\theta }_1,{\theta }_2,k^2)$  is strongly 
peaked at small values of $\vert \Delta\theta \vert $,
 of order $\sqrt{b/A}$. 
 This suggests the  decomposition of ${\cal W}({\theta }_1,{\theta }_2)$
as a sum of two terms involving a local  potential $U_w ( \theta)$
  plus a  small  correction:
\begin{eqnarray}
  {\cal W}({\theta }_1,{\theta }_2) & = & \(( U_w ({\theta }_1)+U_w ({\theta }_2 ) \))/2 +
\Delta {\cal W}({\theta }_1,{\theta }_2) \\
U_w ( \theta) &=&  {\cal W}(\theta ,\theta )=  -{b\over 2 A}  R ( {A \over b} \sin^2 \theta ) 
               \(({1-\cos\theta \over 1+\cos \theta } \))
\end{eqnarray}
A  Taylor  expansion of
$ \Delta {\cal W}({\theta }_1,{\theta }_2) $ with respect to $\Delta\theta$, 
 with  $ \theta $  kept fixed, gives:
$   \Delta {\cal W} \propto {\Delta\theta}^2 \(( 1 +O(  {\Delta\theta}^2) \))$.
It  confirms  that the contribution of $ \Delta {\cal W} $ relative to 
that of ${\cal W}$ is of the order of   $ b/A $. To be   more quantitative,
 we have computed the two averages  $\langle \cal W  \rangle$ and 
 $\langle\Delta {\cal W } \rangle $ using as weight
function $ T({\theta }_1,{\theta }_2)  \sin{\theta }_1 \sin {\theta }_2 $.
Taking $ b= .14 A $ we have obtained 
$ \vert \langle \Delta {\cal W} \rangle / \langle{\cal W}  \rangle \vert=0.013 $.
It appears  then  quite  legitimate to  set $ \Delta {\cal W}  =0 $. 
This approximation allows us to write the  partial derivative of 
the transfer operator $\hat{ T }$  as a symmetrized operator product:
\be
{\partial\hat{ T }(\alpha,k^2)\over  \partial\, k^2 } =
{ \(( \hat{ T }(\alpha,k^2) \, {\hat{U}}_w +
{\hat{U}}_w \,\hat{ T }(\alpha,k^2)\)) \over 2}
\ee
where $ {\hat{U}}_w $ stands for the operator 
associated with  the local potential   $ U_w ( \theta) $.

 Introducing this formula in eq. (\ref{depsk2}) and
remembering that  $ \vert t_0 \rangle $ is an eigenstate  of  $\hat{ T }$,
  the partial derivative 
 $ \epsilon_0(\alpha,k^2) $ takes the simple form: 
\be
  { \partial  \epsilon_0(\alpha,k^2)\over \partial\, k^2 }  = 
{ -A \,  \langle t_0 \vert \hat{ T } \, {\hat{U}}_w+
 {\hat{U}}_w \,\hat{ T } \vert t_0) \rangle \over 2 \, t_0(\alpha,k^2) }= 
  {1 \over 2 } 
\langle t_0 \vert \(({1-\cos\theta \over 1+\cos \theta } \))
            R (  {A \over b} \sin^2 \theta )  \vert t_0 \rangle  
\label{deregur}
\ee

The above result   provides     a very clear evidence that
 the discretization of the RLC model generates  the angular cutoff
 around $\theta = \pi$ which we needed.
Indeed, the  integral involved in the above quantum average is now well defined 
for $ \theta= \pi$, due to the presence of 
the cutoff function $ R (  {A \over b} \sin^2 \theta ) $. 

 Formula  (\ref{deregur}) leads
in a natural way, to a formulation of  a regularized version of the continuous
RLC model. 
The regularized operator  Hamiltonian $ {\hat H}_{RLC}^{r} $
is obtained from the Hamiltonian $ {\hat H}_{RLC} $ given in
equation (\ref{HRLC}) by multiplying the singular " writhe " potential by the 
cutoff function $ R (  {A \over b} \sin^2 \theta ) $.
Standard Quantum Mechanics rules applied to $ {\hat H}_{RLC}^{r} $ 
 give for   $\partial_{k^2} {\epsilon}_0$ 
a formula identical to (\ref{deregur}), provided it is legitimate to neglect the small
difference between  $\vert t_0 \rangle $  and  $\vert \Psi_0 \rangle $, 
respectively eigenstate of $ \hat{ T } $ and $ {\hat H}_{RLC}^{r}(k) $.
We have verified that it is indeed so by comparing the results obtained 
from the above regularized RLC continuous model and those given by 
the transfer matrix method: they do coincide to the level of few $ \% $,
at least in the domain of stretching force explored in the present paper. 
\begin{figure}
\centerline{\epsfxsize=10cm
\epsfysize=10cm
\epsffile{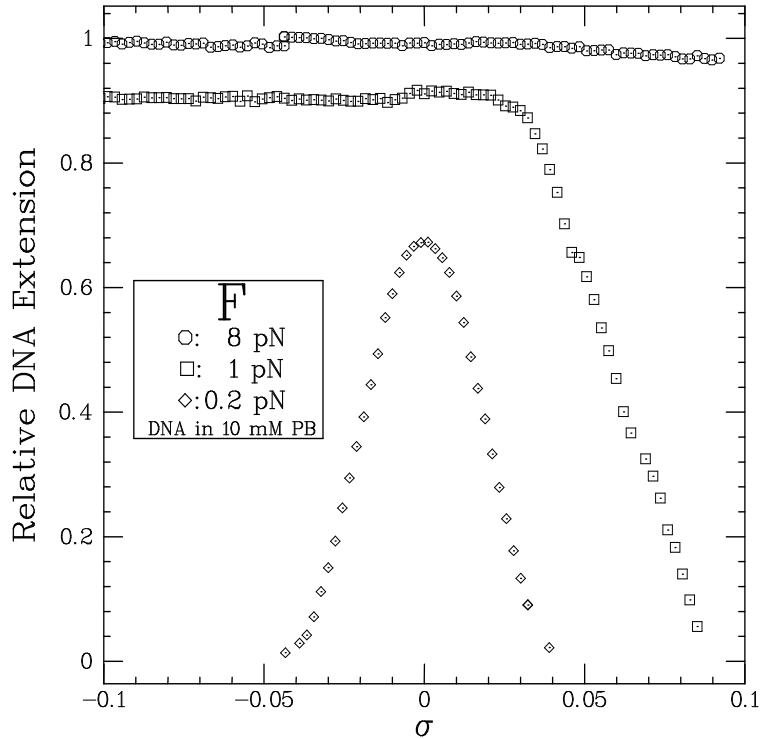}
}
\caption{The  elongation versus  the reduced supercoiling parameter
 $  \sigma ={ n \over L_{k0}} $
 where $  L_{k0} $ stands for the number of double helix turns
 in absence of external constraints. The curve $ F=0.2 \, pN $, a typical 
"hat" curve, corresponds to the regime of entropic elasticity which is well
described by the RLC model introduced in the present paper. The curve $ F=1 \, pN $  exhibits a 
plateau in the underwound region ($  \sigma  < 0 $) which is associated
with the denaturation of the DNA. In the third curve $ F=8 \, pN $   a second plateau
has developed for  ($  \sigma  > 0 $); it has been interpreted as an  induced 
transition to a new structure of DNA: the "P-DNA".
 }
\label{fig1}
\end{figure}
\section {The RLC Model for a quantitative analysis  of supercoiled DNA 
stretching  experiments}
\label{quant_ana}

In this section we give the basic tools allowing to compute,
within the RLC model,
 the various quantities
measured experimentally.

We shall first study the "hat" curve which
is a graph giving, for a given force $ F$,
 the relative extension  of molecule 
 $ \langle { z(L)\over L} \rangle $ along  the stretching force 
 $ \vF $, as a function of the supercoiling angle 
 $ \chi= 2 \pi n $.
  Figure \ref{fig1} gives three examples of such "hat" curves taken 
 within a  rather large 
 range of forces. The supercoiling angle is   parameterized  here
 by the ratio $  \sigma ={ n \over L_{k0}} $
 where $  L_{k0} $ stands for the number of double helix turns
 in absence of external constraints.
  Positive values of $ \sigma$  correspond to  overwinding, negative
  values correspond to 
underwinding.
 As it is apparent in  Figure \ref{fig1}, only the curve associated with 
 a force of $0.2 \, pN$ looks really like a "hat". 
That is so for all  the   extension versus supercoiling curves taken in  the force
range $ .06 \, pN \leq F \leq 0.45 \, pN $. In contrast, when $ F  \ge .5 \, pN $ a 
plateau  develops  in the underwinding region. 
This suggests that the  torsional energy  is 
  converted into  chemical
energy instead of entropic elasticity energy. 
 More precisely,  convincing experimental
arguments  have been given \cite{strick2,strick3} in favour of  the following  mechanism:
 the underwinding  energy is used by the molecular system 
 to open the  hydrogen bonds linking the bases and, as a consequence,
denaturation bubbles appear along the DNA molecule.
  When the force is further increased, say above
 $ 3 \, pN$, a plateau appears  also in the overwinding  region. This has
 received
 a very interesting interpretation \cite{pDNA} as 
a transition towards a new structure of the DNA:
 the so called  " DNA-P ", which is  also predicted by  numerical simulations.

  In the present paper, we shall focus 
 our analysis on  the true " hat " curves,
 {\it i.e } those symmetric under the exchange $ \sigma \Leftrightarrow  -\sigma 
$, 
 as observed for  $ F \leq 0.45 \, pN $.

   \subsection{ The saddle point method}

The experiments suggest that the variations of the relative elongation
  $ \langle { z(L)\over L} \rangle $  versus the supercoiling angle $ \chi$ 
scale as a function of  $\chi/L$. It is convenient 
to introduce  an intensive supercoiling variable
$ \eta = \chi A/L $  which will  turn out to be at most of the  order of few 
units  in the 
domain we are going to explore. It is related to the usual variable $  \sigma 
$
by :
 \be 
         \eta= { \chi A \over L }= { 2 \pi A \over p} \,\sigma =94.8 \,\sigma
\label{etasig}
 \ee
where we have used the values for the pitch: $ p=3.4 \, nm $ and the persistence 
length:
 $ A=51.3  \,nm $.

Let us  rewrite  the partition function $  Z(\chi,F) $  given by equation 
(\ref{Zchi}) 
in terms of scaling variables:
\be
Z ( \eta,\alpha ) \simeq \int dk \  \exp - \frac{L}{A} \((  
{\epsilon}_0(k^2,\alpha) +
{k^2  A \over 2 C} -i \eta \,k  \))   
\label{Zeta}
\ee
The  above partition function  can be computed by the saddle point
method in the limit
$ L/A\gg 1$ with $ \eta $ kept fixed. The saddle point is imaginary, $k_c=i
\kappa(\alpha)$ and is given by  the equation:

\be
{A \over C} +2 {\partial \eps_0 \over \partial k^2}(\alpha,-\kappa^2)={\eta
\over \kappa} 
\label{etakappa}
\ee
The saddle point contribution to the partition function $ Z ( \eta,\alpha ) $
reads as follows :
\be
\ln \(( Z ( \eta,\alpha ) \)) = -{ L\over A} \((  \eps_0(\alpha,-\kappa^2)-
{\kappa^2 A\over 2 \,C }+ \eta \, \kappa  \))+ O(1)   
\label{Zsad}                                        
\ee 
 Let us first compute the torque $ \Gamma $ acting upon 
the free end of the molecule. 
 The experiments to be analyse in this paper were not designed to measure 
$ \Gamma $  but there is  an  experimental project at ENS-Paris  aiming
at its direct empirical determination.   
 \bea
 { \Gamma \over k_{B} \,T}  & = -& {\partial \ln Z \over \partial \chi}=
-{A \over L}{\partial \ln Z \over \partial \eta}   \nonumber \\
  & = &  \kappa + { \partial \kappa \over \partial \eta  }\(( \eta - {\kappa 
\,A\over C}-
 2\, \kappa {\partial \eps_0 \over  \partial k^2 }(\alpha,-\kappa^2) \)) =  
\kappa 
\label{torq}         
 \eea 
where the term proportional to  $ { \partial \kappa \over \partial  \eta}  $ 
vanishes because of the saddle point  equation (\ref{etakappa}).  
Therefore we have found that $\kappa=\Im (\,k_c) $
  is equal to the torque   $\Gamma$ in units of $ k_{B} \,T$.
One can introduce  the thermodynamic potential 
${{\cal G} \over k_B T }= -\ln \(( Z ( \eta,u ) \))- \kappa  \chi $ and  verify 
  that the  supercoiling angle $\chi $ given  by equation (\ref{etakappa})
satisfies  the thermodynamic relation:
 $ \chi= - {\partial \over \partial \kappa } 
\left( {{\cal G} \over k_B T } \right)$.

      We are now going to compute  in the same way the relative molecule 
elongation 
  $ \langle { z(L)\over L} \rangle $ :
\be   
 \frac{ \langle \,  z(L) \, \rangle }{L} =   
\frac{k_B T}{L}\frac{\partial\ln \,Z }{ \partial F}=
-\frac{\partial {\epsilon}_0(\alpha,-\kappa^2) }{ \partial \alpha}
\label{extens}
\ee
As in the computation  of the torque the term proportional to  
$ { \partial \kappa \over \partial  \alpha }  $  does not appear  because it is 
multiplied
 by a factor which vanishes at the saddle point.  In 
other words,
the elongation is given by the same expression whether the experiment is 
performed
under the conditions of fixed surpecoiling angle $ \chi $ or fixed torque $ 
\Gamma $.
In contrast, the situation is very different for the elongation fluctuations:
 $  \langle \,  z(L)^2 \, \rangle - {\langle \,  z(L) \, \rangle} ^2 $ 
 turns out to be  much larger if measured
when the torque is kept fixed instead of the supercoiling angle, as we shall discuss
later.

 {\it From the knowledge of $\epsilon_0(\alpha, -\kappa^2)$,
 using  jointly equations (\ref{etakappa}) and 
(\ref{extens}) one
gets a parametric  representation  of  the  " hat " curves, the parameter being 
the
torque in $ k_B \,T $ unit.}
  \begin{center}
 \subsection{ Solving the  quantum mechanics  eigenvalues problems  
associated with  the RLC model } 
  \end{center} 
In order to complete the description of the theory
 we now sketch the methods used to get  the final theoretical ingredients:
 the groundstate  eigenvalue   $ \epsilon_0(\alpha,k^2 ) $ and its partial 
derivatives.
We have followed two approaches. 

 \subsubsection{ Method a): Ground state eigenvalue of the hamiltonian 
${\hat{H}}_{RLC}^{r}$
associated with  the regularized continuous RLC model}
In the first approach, the computations  are performed within the continuous RLC 
model 
 regularized according to the cutoff prescription derived in section IV.B.
What we have to do  is to solve  the ground state  eigenvalue~problem: 
 ${\hat{H}}_{RLC}^{r} \, {\Psi}_0(\theta)=
{\epsilon}_0\,{ \Psi}_0(\theta)$. This ground state 
wave function is obtained as an  appropriate
 solution of the ordinary differential equation associated 
 with the eigenvalue problem:
\be
	\frac{1}{ 2\,\sin\theta }\frac{\partial}{\partial\,\theta}(\sin\theta \
	\frac{\partial}{\partial \theta}{\Psi}_{0}(\theta))  +
 \(( - V_{r} ( \alpha,-{\kappa }^2 )+ {\epsilon}_0 \)) {\Psi}_{0}(\theta)=0 .
\label{ShRLC}
	\ee  
where  the regularized potential $  V_{r} ( \alpha,k^2 ) $ is given by
\be 
V_{r} ( \alpha,k^2 )= -\alpha \cos\theta +
 {k^2 \over 2} {1-\cos\theta \over 1+\cos\theta} \,
  { I_1(  {A \over b} \sin^2 \theta )\over I_0(  {A \over b} \sin^2 \theta)}
\label{VRLC}
\ee
 We search  for a solution which satisfies regularity conditions both
for $\theta =0 $ and $\theta=\pi$. These requirements are  necessary  
and sufficient to guarantee that the differential operator
${\hat{H}}_{RLC}^{r} $  is a self-adjoint operator.
 They  can  be fulfilled if and only if
the reduced energy $ \epsilon $  belongs to a discrete set of values. 
As a first step, one constructs by  series expansion two solutions of 
(\ref{ShRLC}),
${\Psi}_{a}(\theta) $ and $ {\Psi}_{b}(\theta) $, which are respectively  
regular
for $\theta =0$ and $ \theta = \pi $. One then proceeds to an  outward numerical
integration of ${\Psi}_{a}(\theta)$ and inward numerical integration of $
{\Psi}_{b}(\theta) $ up to an intermediate  value of $ \theta = {\theta}_0 $. 
The
energy eigenvalue equation  is obtained as a  matching condition for the two 
wave functions,
which ensures the regularity  of the eigenfunction for the whole physical domain 
of
$\theta$.  One writes, for $ \theta = {\theta}_0 $, the equality of the 
logarithmic
derivatives of ${\Psi}_{a}(\theta)$ and ${\Psi}_{b}(\theta)$ :
\begin{equation}
	 \frac {\partial{\Psi}_{a}}{\partial \theta}/ {\Psi}_{a} 
	 - \frac {\partial{\Psi}_{b}}{\partial \theta}/ {\Psi}_{b}=0 .
\label{eqvp}
	 \end{equation} 
 Equation (\ref{eqvp}) is then solved by a standard iteration method requiring
a trial approximate eigenvalue. 
The construction of the parametric " hat " curve for a given
value of $\alpha $   begins   with a small value of $\kappa$ ( the
 energy $ {\epsilon}_0 (\alpha ,0) $  is easily  obtained from the results of 
ref. \cite{bouc99} ).
The trial eigenvalue  needed to solve eigenvalue equation (\ref{ShRLC}) for a 
small value of 
$ \kappa $ is easily  obtained  from a first order perturbation calculation.  
 	Since the present method of solving the eigenvalue problem  
automatically
yields the eigenfunction ${ \Psi}_0(\theta)$, the partial derivatives 
of eigenvalues are readily obtained by taking the quantum average 
of the corresponding  partial derivatives of the potential 
$ V_{r} ( \alpha,k^2 ) $.
Let us quote   the derivative with respect to $ \alpha $ ( see  equation 
(\ref{deregur}) 
 of section IV.B. for the derivative with respect to $ k^2 $)
\be  
\frac{ \langle \,  z(L) \, \rangle }{L}= 
                        -\frac{\partial {\epsilon}_0(\alpha,k^2) }{ \partial 
\alpha} =
 \langle  {\Psi}_0\,\vert \cos\theta \vert \, {\Psi}_0\rangle 
\label{cosav} 
\ee
 Once an eigenvalue is known for a value of $ k^2$,
 one  proceeds to the neighboring value $ k^2 + \Delta k^2 $,
  by using $ {\epsilon}_0 + \Delta k^2 \, \frac{\partial 
{\epsilon}_0}{\partial\, k^2} $ 
as a trial eigenvalue and one proceeds to
cover step by step  the desired range  of $  k^2=-{\kappa }^2 $. 
The fact that the state found with this procedure is really the ground state
can be checked from the fact that the wave function has no node.
 
\subsubsection{Method b): The transfer matrix iteration} 
The search for the smallest 
eigenvalue  $ t _0(\alpha,k^2) $ of the transfer operator  $ \hat{ T }(\alpha,k^2)
 $ defined by eq. (\ref{recur2}), (\ref{mtrans}) and (\ref{besarg})
 is done by iteration of  the mapping 
 $\vert z_{n} \rangle = \hat{ T} \,\vert z_{n-1} \rangle $.
 Indeed,   $ \vert z_{N} \rangle ={\hat{ T }}^N \,\vert z_{0} \rangle = 
\sum_n \vert t_n \rangle \langle t_n \vert  z_0 \rangle \,t_n^N  $ is reduced to 
the lowest 
eigenvalue contribution when $ N \rightarrow \,\infty $. Let us write the  
above mapping 
as a linear functional transform:
\be
 z_{n}(k,{\theta}_n) = \int_0^{\pi}  \sin {\theta}_{n-1} d{\theta}_{n-1}\,
    T({\theta }_n,{\theta }_{n-1} )\,z_{n-1}(k,{\theta}_{n-1})
\label{recurfon}
\ee
In order to perform the  integral over $ {\theta }_{n-1} $   we use the following  
discretization procedure. We divide the variation interval $ 0 \leq {\theta }_{n-1} 
 \leq  \pi $ into 
$ n_s $  segments $ { (s-1)\pi \over n_s } \leq {\theta }_{n-1} \leq  { s\,\pi 
\over n_s } $.
The integral over  each segment is done with   the standard Gauss method  
involving
$ n_g $ abscissas and $ n_g $ attached weights. The integral over the full $ 
{\theta}_{n-1}$
   interval is then  approximated by a discrete weighted  sum over $ d =n_s \, 
n_d $ points:
$$
 \int_0^{\pi}  f( {\theta }_{n-1})\sin {\theta}_{n-1} d{\theta}_{n-1}=
\sum_{i=1}^ {d} w_i \,\sin{\theta}_i f( {\theta}_i )
$$
With the above  integration  procedure  each  iteration  step  can be reduced to  
 a linear mapping in a $ d $ dimension Euclidean space $ {\cal {E}}_d$. To each 
function $ z_n(\theta )$
 is attached a vector $ {\vZ}_n $ with $d$ components
 $  ({\vZ}_n)_i= \sqrt{ w_i\, \sin {\theta}_i }\, z_n( {\theta }_i ) $.
The additional factor  has been introduced 
in such a way that the Hermitian norm $ \langle z_n \vert  z_n \rangle $ 
(defined from  our discretized integration procedure) coincides with the 
Euclidian norm $ {\vZ}_n\cdot {\vZ}_n  $. The  linear mapping  in $ {\cal{E}}_d$
is written as: 
$   {\vZ}_n  = {\cal T } \, {\vZ}_{n-1}  $, where the  elements  of 
symmetric  $ d \times d $ matrix  $ {\cal T } $ are given by:
$$  {( {\cal T } )}_{i , j} = \sqrt {w_i \, \sin {\theta}_i \,w_j \, \sin 
{\theta}_j  } \,
T(\alpha,k^2,{\theta }_i,{\theta }_j) $$
It is also easily verified that within our finite sum integration
procedure  we have: $
 \langle z_n \vert \hat{ T}( \alpha,k^2) \vert z_{n-1} \rangle =
                    {\vZ}_n \cdot {\cal T}{\vZ}_{n-1} $.
In order to give a simple criterion of convergence, it is convenient to require 
that
at each iteration step the  vector $  {\vZ}_n $  stay on the $ {\cal {E}}_d$ 
unit  sphere. This is  achieved  by using  the following non linear mapping:
 \be 
   {\vZ}_n  ={ {\cal T } \, {\vZ}_{n-1} \over \sqrt{ {\vZ}_{n-1} \cdot {\cal 
T}^2  {\vZ}_{n-1} } }
\label{receucl}
\ee     
 The criterion   of convergence   is  set up  as follows :

Let  us define  $ \Delta  {\vZ}_n={ \vZ}_n-{\vZ}_{n -\Delta n }$ where $ \Delta n $ 
is an integer, 
 typically of the order  of one hundred. ${ \vZ}_n $ is considered to be an 
appropriate fixed 
point ${\vZ}_f$ if $ \sqrt { \Delta  {\vZ}_n \cdot \Delta  {\vZ}_n } \leq \delta 
$.	In practice we
have taken $ \delta =10^{-4} $;  this choice , together with  $ \Delta n= 200 $,  
leads to a very 
tight convergence test.

Knowing   ${\vZ}_f$   it is a straightforward matter to get the groundstate 
energy 
$ {\epsilon}_0(\alpha,k^2)$ and its partial derivative with respect to $  \alpha 
$ and
$ k^2$:
\bea
{\epsilon}_0(\alpha,k^2) &=& -{A\over b}\ln t_0(\alpha,k^2)=
- {A\over b}\ln\((  {\vZ}_f \cdot  {\cal T } \,{\vZ}_f \)) \nonumber \\
{ \partial \epsilon_0(\alpha,k^2) \over \partial\, k^2  } & =
&- {A\over b \,t_0(\alpha, k^2) }\,  {\vZ}_f \cdot { \partial {\cal T }\over 
\partial\, k^2  } 
 \,{\vZ}_f  \nonumber\\
{ \partial \epsilon_0 (\alpha,k^2 ) \over \partial\, \alpha  } & =  
& - {A\over b \,t_0(\alpha, k^2) }\,  {\vZ}_f \cdot
 { \partial {\cal T }\over \partial\, \alpha  }  {\vZ}_{f}
={\vZ}_{f} \cdot {\cal C }os \,{\vZ}_{f}
\label{formtrans}
\eea
where the ${\cal C }os$  matrix  elements are given by:
 $ {\(({\cal C }os\))}_{i,j}= {\delta}_{i,j}
      \cos{\theta}_i  $.

 It is also possible to get an explicit expression of the eigenfunction $ {\Psi}_0 ( 
\theta) $ from the 
knowledge of the fixed point vector ${\vZ}_{f}$. The eigenfunction obeys the 
integral equation:
\be
 {\Psi}_0 ( \theta) = 
{1\over t_0}\int_0^{\pi} T(\theta, {\theta}_1)
 {\Psi}_0 ( {\theta}_1) \sin {\theta}_1 d \,{\theta}_1 
\label{wfinteq}
\ee
Using our finite sum method to perform the  integral over ${\theta}_1$
in   the r.h.s. of the above equation,  $ {\Psi}_0 ( \theta) $ is obtained 
immediately 
in terms of the components $ Z_{f,i}$ of the fixed point vector:
$$ 
 {\Psi}_0 ( \theta) ={1\over t_0} \sum_i T(\theta, {\theta}_i) 
\sqrt{ w_i\, \sin {\theta}_i } \, Z_{f,i}
$$

In our implementation of the method
we have taken $ n_g =10$ . The use of higher values of $n_g$ may 
lead to overfitting.

In contrast we can
vary more  freely the number of sectors $ n_s$. 
The ground state energy  ${\epsilon}_0(\alpha,k^2)$  and its
partial derivatives vary by less than one part per million
  when $ n_s $ goes from 2  to 8 with $ n_g =10 $. 
One has to go down to $n_g=8$ and  $ n_s=2$ to see a variation of about $ 
10^{-5}$. 
We have   verified, using  the high precision integration subroutines  
provided by 
the Mathematica software,  that  the above wave function satisfies, in typical 
cases,
 the  eigenvalue  integral equation (\ref{wfinteq}) to better than $ 10^{-9}$  
with the choice 
$ n_s=4$ and $ n_g=10 $.

The vector  $ \Delta  {\vZ}_n={ \vZ}_n-{\vZ}_{n -\Delta n }$  introduced to set up
the convergence criterion  can be used to get the 
energy $ {\eps}_1 $ and the eigenfunction $ {\Psi}_1 ( \theta) $  of the
 first excited state  with a fairly 
good accuracy, say better than  $ 10^{-3} $.
In particular the variation of the energy gap $ \Delta \eps=    {\eps}_1 - {\eps}_0 $
versus $ \eta $  will shed some light upon the crossover
phenomenon to be discussed in section \ref{tw_wr_pl}.
\begin{center}
 \section{ Monte
 Carlo Simulations with the Discretized RLC model}
 \label{Monte_Carlo}
\end{center} 
The Monte Carlo procedure allows to generate configurations of the discretized
RLC with a frequency proportional to their Boltzmann weight. A full simulation 
of the discrete model, incorporating the self avoidance effects, the
check that only unknotted configurations are kept, and the estimation of the
Writhe with the non local Fuller formula was developed in \cite{volog}
in the case of closed DNA chains.  Vologodskii and Marko \cite{marvol} were the first 
 to adapt this formalism to the case of a  single open  supercoiled chain. In order to
facilitate  the transition from closed chain  to open chain, they assume that the
chain is confined  between two impenetrable walls, with  the two free ends sitting 
on different walls. From an experimental point of view, one wall is certainly welcome 
since in the actual experiments one molecule end is achored by biochemical
links to a glass plate. The other end
 is  biochemically glued  to a magnetic bead. In the experiments studied in this paper,
 the bead radius is one seventh of the molecule contour length  L,
 so that an unpenetrable wall
looks  somewhat  inadequate to account for  
the geometrical obstruction  of  the magnetic bead, even if one neglects 
the very slow processes where the molecule releases the supercoiling 
by turning around the bead.
The authors were, of course, aware of the problem and thus they limited their comparison to
experiment to relative extension larger than $ 0.3 $. This  wall effect is  
visible in the limit of zero supercoiling limit where the M.C. results differs 
  significantly from the WLC at forces below $.1 pN$ ( in that case 
 finite size effects may also play a role since $L/A \sim 10$ to $20$).
If the two conditions $ < z >/L > 0.3 $ and $ F > 0.1 pN  $ are satisfied then 
 a reasonable agreement was achieved  with 
experiments over a rather broad range of  forces and supercoiling.
The agreement is particularly satisfactory for 
force extension curve at $ \sigma =0.031 $  and this 
 has allowed the authors to give an  estimate of the 
twist rigidity $ C $, with a $ 20 \% $ error bar . 
We shall return to this last point in section VII. 
Vologodskii and Marko have also investigated the effect of  the reduced ioning 
strength which governs the DNA effective Coulomb radius, which
can be characterized by the Debye length $\lambda_D$.
 Passing from 20 mM  to 200 mM of NaCl  leads
 to a Debye length  reduction  by  factor  $\sqrt{10}$.
The calculated  changes in the force extension curves
 at $ \sigma= 0.01 $ and $  \sigma= 0.03  $
are barely visible for  stretching forces
 in the range  $ 0.1 pN \leq F \leq 0.4 pN $. This includes 
 the  $ (F,\sigma )$ domain  to be considered in the present  paper. 
In the actual  experiments,  the  Debye length  corresponds to
 30 mM  of NaCl.  Therefore the Vologodskii and Marko \cite{marvol} findings
suggest that the self  avoidance effects associated with the finite 
Coulomb radius 
play a relative minor role in the data 
to be analysed in the next section.

They  present  some evidence for  the importance  
of knotting supression in  the particular case  $ F=0.2 \,pN $ and 
$ \sigma =0.05 $. A  typical simulated  knotted configuration 
is seen to have  the ability to absorb the supercoiling more efficiently than
  the corresponding unknotted one: 
it leaves the chain with a larger extension. 
Our  analysis of the $ F=0.2 \, pN $ hat curve  of Fig \ref{fig1} 
will concern the range $ \vert \sigma  \vert \leq 0.016 $ which lies
far away from the  borderline  point considered by the authors.

In the work  presented here, we have performed a much less ambitious Monte
Carlo simulation. 
  We kept within the RLC model without self-avoidance, and the simulation
was used essentially as a guide to validate our model, using the local formula for $\chi_W$ 
together with a discrete version of the chain: we checked that it is 
free of pathology and able
to reproduce the experimental findings. 
It suggested to us  that the length  cutoff $ b$ provided  by the elementary link of
the discretized chain could generate the angular cutoff needed to 
regularize the continuous RLC model, 
as it is shown  in section IV.B. Our first computations 
within the regularized continuous model ( the method a) of subsection V.B.1 ) were 
 checked on few points by Monte-Carlo simulations and the agreement gave us confidence 
in  our approach. 

Subsequently  we gave
our preference to the transfer matrix
iteration method of subsection V.B.2  which uses exactly the same
theoretical inputs as the Monte Carlo simulation  but  leads to accurate
results with much less computer time.       
The fully discrete model involving three Euler angles per elementary link can
be simulated, but it is more efficient for our purpose to make use of
the fact that the $\psi$ integrals can be done analytically, and thus to work 
only
with the two angles $\theta$ and $\phi$ for each link. Indeed, using our
computation of section 3 and integrating over the momentum
$k$ conjugate to the supercoiling angle $\chi$ after integrating out the $\psi$ 
angles,
 the partition function of the supercoiled
DNA molecule can be written as:
\be
Z(\chi,F)= \int {\cal D} (\theta,\phi) \exp\((- {E_{WLC} \over k_B T}
-{C \over 2 L} \(( \chi-\int ds \; \dot \phi(1-\cos \theta) \))^2 \)) 
\ee
This is the path integral which we have discretized and simulated.
The discretization procedure is exactly the one described in the previous 
section.
We use $N$ elementary links of length $b$, 
and the two discretization rules (\ref{rule1},\ref{rule2}) for the energy
function. The path integral measure is substituted by 
$\prod_{n=1}^N d \phi_n d \cos \theta_n$. The partition function is thus
expressed as a $2 N$ dimensional integral, and the corresponding probability
measure can be sampled by Monte Carlo.

We use the standard Metropolis algorithm, where a new configuration of the chain 
is proposed
at each step, the corresponding change of energy $\delta E$
is computed, and the change is 
 accepted with probability $min(1,\exp(-\delta E/k_BT))$.
The point on which one must be careful is the choice of moves. Clearly a choice
of local moves (changing $\theta_n,\phi_n$ on one link at a time) is a very bad
one with which a macroscopic change of the molecule takes very many steps,
and it leads to very long thermalization times. We have checked that it is
in effect useless. We need to implement more sizeable moves, but moves
which do not change the energy too much. One method which we have found rather
 effective is the following. We have first relaxed the constraint involving
$\theta_N$,
 which should not change the extensive properties of the chain. Then the moves 
 consist in:
 \begin{enumerate}
 \item One picks up one link of the chain, number $n$.
 \item One rotates the section of the chain $j\in {n+1,...,N}$ by a global 
rotation.
 The rotation, of angle $\gamma$ around an
 axis $\vec n$, is picked up at random, with a uniform distribution for
 the choice of $\vec n$ on the unit sphere, and a uniform distribution
 of $\gamma$ on an interval $\lbrack -\delta,\delta \rbrack $ (a value of $\delta$ of order .5
 is generally adequate).
 \item One moves to the next link and iterates the procedure.
 \end{enumerate}
This is one of the types of moves which are used in the standard 
simulations of supercoiled DNA \cite{volog}.

We have simulated mostly chains of $300$ links, each link
$b$ having a length of one
 tenth of the persistence length. We have checked that 
the
finite size effects can be neglected with respect to the statistical errors
for this value. For a given value of the supercoiling angle $\chi$
(or rather of its intensive version $\eta$), we perform a simulation,
with a number of Monte Carlo steps per link of order $10^4$. The first third of 
data is used for thermalization, the rest is used for measuring the distribution
of elongation. 

We show in figure \ref{fig3}  the  results for $<z>/L$, and the statistical error bars.
The computations were done at $C/A=1.4$, and $b/A=.10$
for values of the force  equal to $.116 \, pN$, and the amount of
supercoiling given by $\eta=0.0, \,0.46, \,0.92, \,1.39$. 
We see that they are in rather  good agreement with the experiment and the  
 transfer matrix results.  
 Note that since we do not have to use the 
two impenetrable walls trick, our computation 
 is also valid in the
 zero supercoiling limit.  

When one increases the degree of supercoiling the thermalization time
becomes prohibitive, and an examination of the configurations
shows that they start building up some small fluctuating plectoneme-like
structures. Clearly in order to study this regime one must first include self
avoidance, and then incorporate moves which are able to shift the plectonemes 
positions  efficiently, as was done in 
\cite{volog,marvol}. We believe that this type of approach
 can be pursued further in order to get 
a precise comparison with experiments in the strongly supercoiled regime. It is 
complementary to the more restrictive, but more analytical, study at small supercoiling
which we  develop here.

\begin{center}
 \section{  Analysis of the experimental data on DNA extension versus 
supercoiling 
 in the small force regime.
}
\label{ana_exp}
\end{center}
In this section we are going to analyse a  limited set   of data obtained by the 
 LPS-ENS group  on the same single DNA molecule. 
They consist of three extension versus supercoiling curves 
with values of the stretching forces
$F= 0.116\, pN , 0.197 \, pN $ and $ 0.328 \, pN $. For these data 
 a point-by-point evaluation  of the systematic uncertainties
is still lacking,  so 
only the  statistical uncertainties 
have  been  considered in the determination of the ratio $ { C\over A} $.
A simplified version of this analysis  has been already 
given by us in a previous publication \cite{BM1}.

We have also performed two cuts upon the data 
in order to exclude from the analysis regions
where   the  validity of   RLC model is questionable.
 The first cut allows to neglect the effect
of the plane onto which the DNA is attached. The second one allows
to neglect  the self avoidance of the chain.

First we exclude values of the relative elongation such that 
$ {\langle  z(L) \rangle  \over L}\leq 0.1 $. Experimentally, one end of the 
molecule is attached to a plane which thus implements a constraint $z(s) >0$,
for the whole chain.
  We shall see later on that when the reduced 
supercoiling
parameter $ \eta$ increases from 0 to few units the probability distribution of $ 
\theta $ 
develops a peak near $ \theta= \pi $.  Specially  when 
$ 0\leq {\langle  z(L) \rangle  \over L} \leq 0.1 $ , it means that
 the RLC model  is likely to generate 
configurations  with  $ { z(s)   \over L} \leq 0$. We think that, for the 
experimental lengths $L$ under study, the regime where
 $ {\langle  z(L) \rangle  \over L} > 0.1 $ is such that typically, the whole 
chain is in the half plane $z>0$.

The second cut  excludes the too high values of the reduced supercoiling angle by 
imposing the condition $ \eta  \leq 1.5 $  for $ F=0.197 pN  $ and $ F=0.328 pN$.  
In a following section we shall present evidence that  when $ F= 0.328 pN $ 
the RLC model generates plectoneme-like configurations 
 above a critical value  $ { \eta}_c \approx1$ .
 Our model ignores self avoiding effects which are
present under the actual experimental working  conditions   since the DNA  
Coulombic potential is
only partially  screened. The experimental plectonemes must   have a radius 
\footnote{The plectoneme' radius is the common radius of the two  interwound superhelices}
larger than  the
DNA Coulombic radius. In contrast,  in the RLC model  plectonemes with an 
arbitrary small  radius can be generated.
  For  a  fixed  variation
  of the  supercoiling  angle, the creation of a plectoneme, having
 a given length, absorbs, on the average   a   smaller
fraction of the DNA chain,   in comparison  with   a situation where self 
avoiding 
 effects are involved. Indeed we have found that when the above constraints  are 
not  fulfilled  the  experimental points have a tendency to fall inside the 
theoretical hat curves, {\it i.e }  for a given $ \eta $, experiment gives a 
smaller
 $ {\langle  z(L) \rangle \over L} $. 
\begin{figure}
\centerline{\epsfxsize=12cm
\epsfysize=16cm
\epsffile{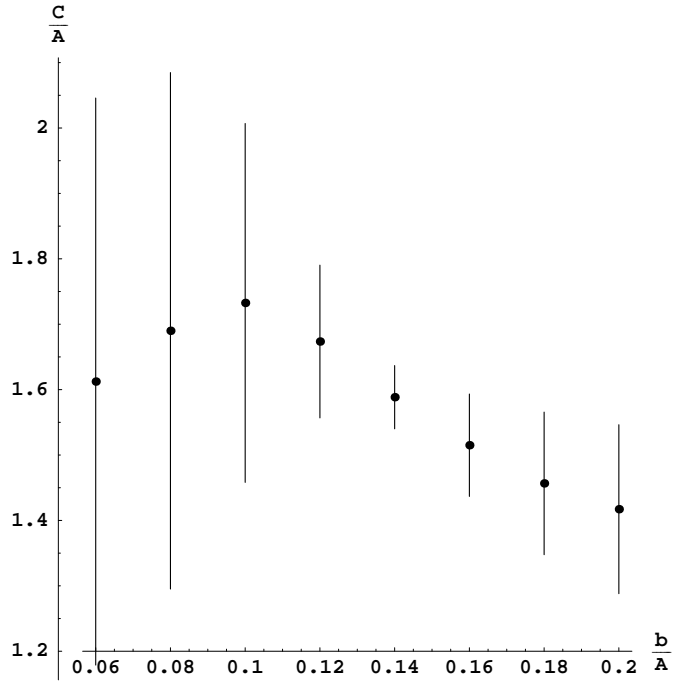}
}
\vspace{-4 cm}
\caption{Empirical determination of  the cut off length b and the ratio $ C/A $
 from the hat curves analysis. In the case $ F=.116 \, pN $,
 we have plotted,  versus the cutoff $b$,
the mean ratio  $  \langle   C/A \rangle $ obtained by averaging
 the  $ C/A$  values   predicted  by  the RLC model from each empirical hat curve point. The error
bar represents for each $b$ the variance 
$ {\sigma}_r $ which leads to a measure 
of  the RLC model ability  to reproduce the hat curve data. 
A remarkable agreement is achieved with $ b= .14 \,A$ while it becomes very  
poor for $ b=.06 \,A $. This is consistent with the RLC 
model singularity near $ b=0$.   
 }
\label{fig2}
\end{figure}

In order to extract the values of $C/A$ and of the cutoff $b$ from the
experimental data, we have used the following technique.
Using equation (\ref{etakappa})  the ratio $ C/A $ can be written as a 
function of the
reduced supercoiling angle $ \eta$  and the torque $ \kappa $ :
 \be
{A \over C} =  {\eta \over \kappa} -2 {\partial \eps_0 \over \partial 
k^2}(\alpha,-\kappa^2)
\label{AoverC}
\ee
 With the help of  interpolation techniques, one can invert equation 
(\ref{extens})  in order
to get $ \kappa $ as a function of  $ {\langle  z(L)\rangle\over L} $.  
In this  way, each  " hat " curve point of coordinates
 $  ( \eta ,\, {\langle  z(L) \rangle \over L} )$ 
is associated  with an empirical value  of the rigidities ratio $ {A\over C}  $,
 once  a choice of the cutoff length $ b $ has been made.  
If the RLC model is to give  a good representation of
the data  there must exist a value of $b $ such that
 the  empirical values $ C/A $ obtained from all points of all
 hat curves  cluster around the correct result for   $ C/A $.
\begin{figure}
\centerline{\epsfxsize=12cm
\epsfysize=16cm 
\epsffile{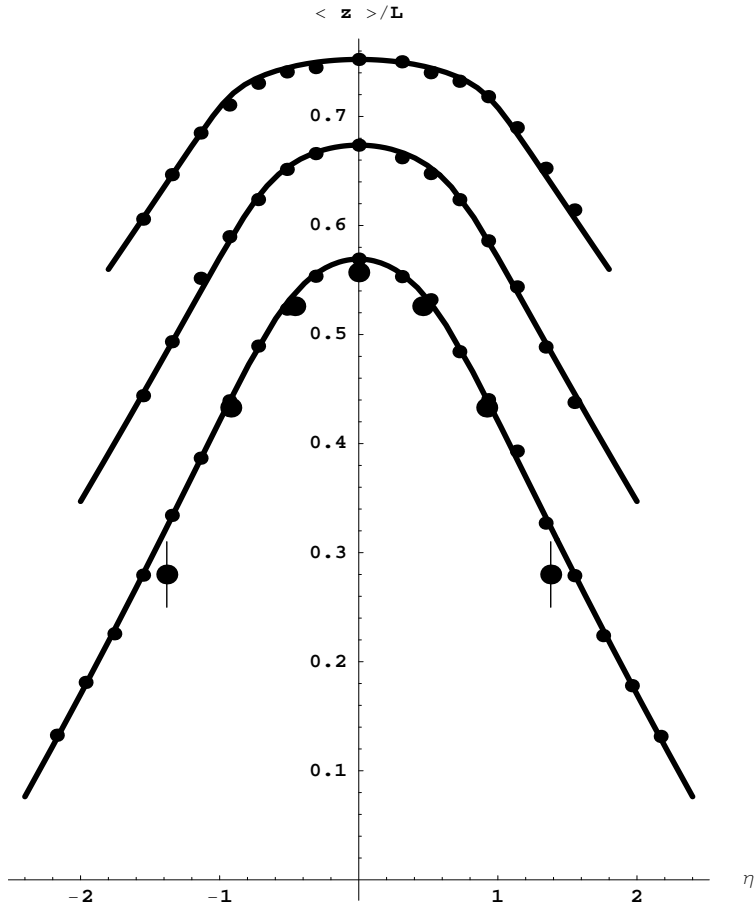}
}
\vspace{-2cm}
\caption{ The  elongation versus reduced supercoiling angle, $\eta =94.8 \,\sigma $,
 for forces $ F=.116, \,.197, \,.328 \, pN $, from bottom to top.
The smaller points are the experimental results, the bigger points on the lowest curve
 are from Monte Carlo simulations and the   full  lines are the predictions 
of the RLC model obtained as indicated in the text. }
\label{fig3}
\end{figure} 

For F =0.116 \, pN, we have plotted on Figure \ref{fig2}   the average 
ratio  $  \langle    C/A  \rangle  $ and the variance 
 $ \sigma_r=\sqrt{ \langle  \, ( C/A  )^2\,\rangle - (\langle  \,  C/A  
\rangle\,)^2  }$
versus the cutoff length $b$. The average and variance are computed over all 
the  20  experimental  points of the  hat curve.
 The point with $ \eta=0 $ is
an input, used to compute the dimensionless force  parameter $\alpha $.  
Using the scaling properties  of the model, we have reduced  
the data analysis to  a two parameters fit: $  b/A $ and $ C/A $. 
Since our model  reduces to the WLC model in the $ \eta=0 $ limit it 
is legitimate to use the persistence length $ A  $  obtained 
within the WLC model  from  the analysis of  the force versus extension
curve \cite{bouc99}. 
As it is apparent on Fig. \ref{fig2}, the  RLC model with $ b=0.14 \, A $
leads to a remarkably good agreement with the data.
It corresponds to  a minimum value of  $0.03 $ of
 the quantity  $ {\sigma_r \over  \langle    C/A  \rangle } $ which
measures the ability of our model to reproduce the data. 
 The best cutoff length  $ b $ value is approximately equal to twice  
 the  double helix pitch.
This is  close to  the length resolution  $\Delta l $, 
which we had to  invoke in order  to justify the assumption of
 a cylindrically symmetric  rigidity tensor. 
It is interesting  also to note  that 
the average value  $  \langle    C/A  \rangle  $  { \it varies slowly with  } $b$.
 The  variance $ \sigma_r  $  increases rapidly if
one goes to  small values of the cut-off length ;
this is consistent with the fact that the RLC model becomes singular in the 
limit $ b \rightarrow 0$.

  Similar   results, somewhat less precise, have been 
obtained for 
the two other values of  $ F $  considered in this section. They favour
 the same value of  $ b/A $ and give values of $  \langle    C/A  \rangle  $
 consistent  with the ones obtained for $ F= .116 \, pN$, with variances
 $ \sigma_r $ about three times larger. 
Performing a  weighted average upon  the whole  set of 
      $ C/A $ empirical values obtained by the three hat curves,
  taking $ b=0.14 \,A$, one gets the following 
empirical determination of the ratio   of the
 two elastic rigidities involved in the RLC model :
\be 
  {C\over A}=  1.64  \pm 0.04 
\ee
This result is in agreement with the value given 
in our previous work \cite{BM1}: $ {C\over A}\simeq 1.68 $.
Since we have used  data  which   do not incorporate in a quantitative way the
systematic uncertainties, the above number should  be considered
 as somewhat preliminary, but  as it stands, it constitutes a
significative  improvement upon the previous empirical estimates.

We give in Figure \ref{fig3} the  theoretical " hat " curves using the ratio
 $  {C \over A }=  \langle    C/A  \rangle $  obtained  from the data at  each force
 and the cutoff  favored value $ b= 0.14 \, A $.  The agreement 
with the experimental data  is very 
satisfactory  and confirms the  overall consistency of the procedure. 
It  should be stressed  that  the rather sharp  bend connecting 
a slow quadratic decrease to a steep  nearly linear falling down, which
is  observed for $ \eta \simeq 1 $ on  the $ F=.33 \, pN$  hat curve, is 
rather well reproduced by our model.

Using the values of the persistence length $A$ obtained from 
 the force versus extension curve  measured on the same single molecule, 
 one can obtain the  following empirical value of the twist rigidity: 
\be 
  C=  84  \pm 10\,nm 
\label{Cempir}
\ee
The statistical error on $ C $  is of the order of 2 $ nm $ but  
the systematic errors are expected to be non negligible and 
 the  overall uncertainty of  $ 10 \, nm $ which  accounts for those
looks reasonable.

It is  here of some interest to quote a recent empirical determination
of the twist rigidity obtained from an analysis which does not 
involve the use of the RLC model or the like. A detailed account of the procedure
is to be found in reference \cite{houch98}. We shall
here only quote the result: $ C=86 \pm 10 \, nm$, which is in good
agreement with the number given by  equation (\ref{Cempir}).

Vologodskii and Marko \cite{marvol}  have given
previously
  an  estimate of $C$  
 obtained from  one  force extension curve at $ \sigma =0.031 $, 
where the agreement with their Monte Carlo simulations
  is particularly striking : $ C=  75  \pm 15 \, nm $.
This result agrees with ours.
Unfortunately they were not able to get any value of $C$, leading to 
an  overall  good fit for  the complete set of  force extension curves.

Moroz and Nelson \cite{mornel1,mornel2} have analyzed the  "hat" curves, using 
a  perturbation  approach to the RLC model. Because of the constraint
$ K= \sqrt{\alpha - {1\over 4} {\kappa}^2 } > 3 $, inherent to their method, they
explore the force  supercoiling angle domain 
 given by $ F \geq .3 \, pN $ and $ \sigma \leq 0.01 $. 
This implies that the ranges of
relative  extension for a given force  are very narrow :  
for the top curve of Fig(\ref{fig3})   ($ F= .33 \, pN $): 
$ 0 .72 \leq   \langle z \rangle/L \leq  0.75 $,
 the higher the force the narrower is  the interval and  
closer to unity is  its center. 
It is clear from Fig \ref{fig3} that the range of relative extension analyzed
 in the present  paper  is much wider. 
In fact there is   a rather small overlap of the  experimental
domains involved in the two analysis :  the intersection of the two data set amounts 
only to 20 $ \% $ of our  present data  set. Moroz and Nelson 
   have incorporated hat curves  data associated with relatively
 high force values: $ 0.6, \,0.8, \,1.3, \,8.0  \,pN $.
 For such forces, the RLC model is certainly
not valid when  $ \sigma <0 $ and even for  $ \sigma >0 $ in the case of the highest force.
They \cite{mornel2} have derived from their two parameter fit $ ( A,C ) $
 the twist rigidity  value $ C=109 \, nm $ (in a preliminary analysis  \cite{mornel1}
based upon a  more limited set of data they gave $ C=120 \, nm $). These authors did not  
give  the uncertainty associated  with $ C $  and  the value of $ A $
coming out from their two parameter  fit.
Because of the limited  range of relative  extension  values they can fit, we believe
that the error bars on their  results should be larger than ours. 
As a comparison, we have used $ C= 109 \, nm $  
to compute the  theoretical hat curves of Fig. \ref{fig3}. As expected,
 the quality of the fit becomes significantly poorer: the overall $\chi_2$  is multiplied
by 5.3 when one  goes from $C=86 \, nm $ to $ C= 109 \,nm $.

\section{Twist, Writhe and Plectonemes.}
\label{tw_wr_pl}    
In this section we would like to use our solution of the discretized 
 RLC model to study, at a given force (F=.33 \, pN), 
the variations of the torque, twist and  writhe  thermal averages versus the
supercoiling reduced angle $ \eta $. As we shall see, there 
exist two  very different regimes.
Below a rather sharply defined value of the supercoiling angle  $ {\eta}_c $,
the DNA chain behaves nearly  as an elastic non-flexible rod. Above  $ {\eta}_c $ 
  the torque becomes nearly independent of  supercoiling while the writhe grows linearly
with it. We shall  give arguments 
suggesting that  in the  high $ \eta $  regime  the supercoiling
 constraint is satisfied by the creation of  plectoneme-like configurations.
\subsection{ Relation between the torque and the average twist }
To begin we are going to prove that the thermal average of the rod twist 
$ \langle \,T_{w} \, \rangle $  is given in terms of the torque $\Gamma$
by   the classical elasticity formula  for  a non-flexible 
rod:
\be
\langle \,T_{w} \, \rangle = { \kappa \, L \over C }= { \Gamma \, L \over \ {\cal C}}
\label{twistav}
\ee
where $ {\cal C}= k_{B} \,T \, C $ is the  usual twist rigidity.
It is convenient to introduce  
the joint probability distribution of a DNA chain configuration $ {\cal P}( \chi_1 ,\chi_2)$
with $ T_W=  \chi_1 $ and $ {\chi}_W= \chi_2 $; it can be written as the double Fourier transform:
$$
{\cal P}( \chi_1 ,\chi_2)\propto Z( \chi_1, \chi_2)= \int { dk_1\,dk_2  \over 4 \pi^2}
\exp(-ik_1 \, \chi_1-ik_2 \, \chi_2 ) \tilde Z(k_1,k_2)
$$
where  $ \tilde Z(k_1,k_2) $ is given by the functional integrals
\begin{eqnarray*}
 \tilde Z(k_1,k_2)  &  = & \int { \cal D }(\theta,\phi) \,\exp  \((
i\,k_2\, \chi_W- {{\cal E}_{bend}+{\cal E}_{stretch}
\over k_{B}T} \)) {\tilde Z}_{T}(k_1)   \\ 
{\tilde Z}_{T}(k_1) & = &  \int { \cal D }(\psi) \,\exp \(( i\, k_1\, T_{w} 
-{{\cal E}_{twist} \over k_B T}\))
 \end{eqnarray*} 
As in section  III B, we perform explicitly the functional integral upon $ \psi $ 
and  get
$ {\tilde Z}_{T}(k_1) = \exp( -{k_1^2L\over 2 C}) $. This leads   
to  the factorization  property: $  \tilde Z(k_1,k_2) =  {\tilde Z}_{T}(k_1)  {\tilde Z}_{W}(k_2 ) $,
where the writhe term  ${\tilde Z}_{W}(k_2 )$  is given by the path integral:
$$ 
  {\tilde Z}_{W}(k)= \int { \cal D }(\theta,\phi) \,
 \exp  \((i\,k\, \chi_W- {E_{WLC}\over k_{B}T} \))
 $$
 { \it As a physical consequence, the twist $  T_{w}$ and the writhe $ \chi_W $ fluctuate 
independently in the RLC model, when no supercoiling  constraint
is applied upon the free end of the molecule. }

The thermal average $ \langle \,T_{w} \, \rangle $,  in presence of the supercoiling
constraint $ \chi= T_{w}+\chi_W $, is  then given by:
\bea
\langle \,T_{w} \, \rangle &=& { 1\over  Z(\chi,F)} \int d\chi_1 \, d\chi_2
\,  \chi_1 \, Z_{T}( \chi_1 )\, Z_{W}( \chi_2 )\, \delta ( \chi_1+ \chi_2-\chi )
 \nonumber \\
 &=&
{1 \over Z(\chi,F)} \int {dk \over 2 \pi}
\((-i {d {\tilde Z}_T \over dk } \)) 
{\tilde Z}_{W}(k) \exp(-i  k\, \chi ) \nonumber 
\eea
Using the explicit form of $\tilde Z_T$,
  one gets  the announced   formula  and by subtraction the writhe 
thermal average :
\bea
\langle \,T_{w} \, \rangle &= &- {L \over C} {\partial \ln Z \over \partial \chi}=
 {L \over C} \, { \Gamma \over  k_{B} \,T }  \nonumber \\
\langle \,\chi_{W} \, \rangle &=& \chi- \langle \,T_{w} \, \rangle =
{2\,L\,\kappa \over A } \, { \partial \epsilon_0(\alpha,-\kappa^2) \over
\partial\, k^2 } 
\eea
\subsection{The average writhe in the zero stretching force limit: a comparison
with  Monte Carlo simulations of self-avoiding closed supercoiled  DNA chains }

\begin{figure}
\vspace{-2cm}
\centerline{\epsfxsize=12cm
\epsfysize=16cm
\epsffile{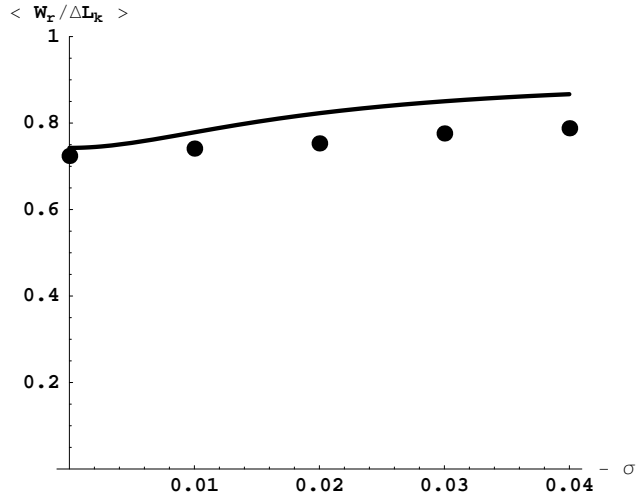}
 }
\vspace{-6cm}
\caption{Comparison of Monte Carlo  ({\Large $\bullet$})  and 
the RLC model ( continuous line)  results for  the reduced average 
$ < Wr/\Delta Lk > = \lim_{F=0}\,  \langle \,\chi_{W} \, \rangle / \chi$. } 
\label{figcompvog}
\end{figure}

The above  formulas offer the opportunity to compare, on a specific 
point, the  RLC model predictions
with  closed DNA chain  Monte-Carlo simulations \cite{volog}, which incorporate  
self-avoiding effects.
Using  the same ratio $ C/A=1.5 $ and the same value of $ b/A=.2 $  as 
Vologodskii et al. \cite{volog},
we have computed
 $ \lim_{F=0}\,  \langle \,\chi_{W} \, \rangle / \chi =< Wr/\Delta Lk > $ 
 for $\sigma $ values
 taken in the range: $ 0 \le -\sigma \le 0.04 $. 
 The  comparison with the  numbers taken from   reference 
\cite{volog} is displayed in  Figure \ref{figcompvog}.
It is apparent that the RLC model computations, and Monte-Carlo simulations are 
in rather good agreement: when $ \vert \sigma \vert \leq 0.02 $  -
 the range   explored in our previous data analysis -
  the results  diverge by less than $ 8.5 \% $ and the difference reaches 
$ 10 \% $ when  $ \vert \sigma \vert  \rightarrow 0.04 $.
 They both agree rather well  with the
measurements performed in references \cite{boles} and \cite{adrian}. 
If we forget about the possible finite  size effects  in 
 the Monte-Carlo simulations where  $A/L \simeq 1/12$,
 the small  deviation may tentatively  be attributed to the
self-avoiding  effects not incorporated in our computations. 
\subsection{ A torque and writhe versus supercoiling cross-over : a possible sign
for  thermal   excitation of  plectoneme-like configurations}
\begin{figure}
\centerline{\epsfxsize=14cm
\epsfysize=18cm
\epsffile{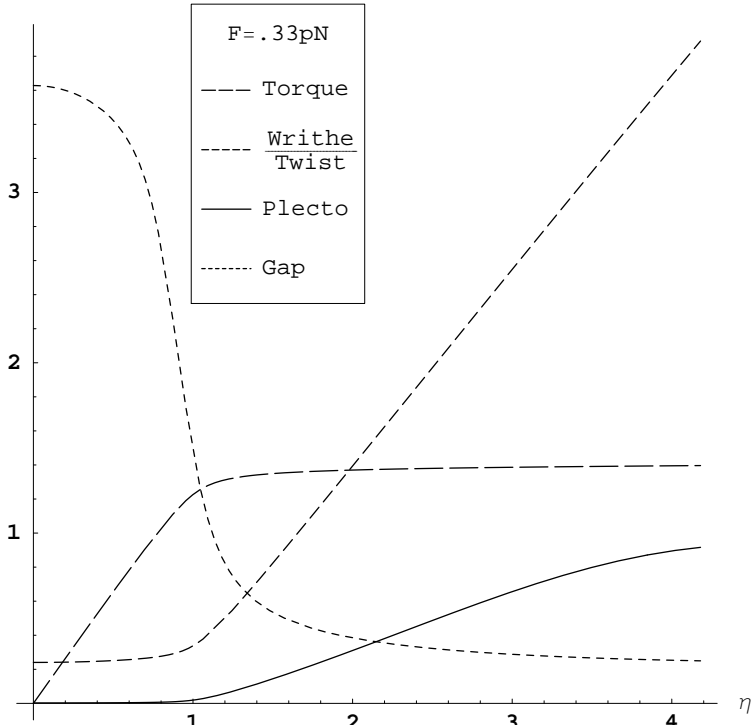}
 }
\vspace{-6cm}
\caption{Twist, Writhe and Plectonemes in the RLC Model. We display several curves
 which indicate 
that a cross-over phenomena is taking place near $ \eta_c \approx 1$. After a linear
increase, expected for a non-flexible rod, the torque becomes independent of the 
supercoiling above $ \eta_c $. Simultaneously, the writhe to twist ratio, which was staying 
constant around $ .2$, starts a rather steep linear increase.
 Near $ \eta_c $ the function $ plecto(\eta)$, 
   which measures the forward backward symmetry of the $\vt$ angular distribution 
relative to $ \vF$, takes off from 0 to the value 0.9.( $ plecto=1$
 for a plectonemic configuration). All these features suggest that 
the cross-over could be attributed to the creation of plectoneme-like configurations. }   
\label{figplecto}
\end{figure}

In Figure \ref{figplecto} we have plotted ( for the case $ F=.33 \, p N $) the torque 
in $ k_{B} \,T $ unit, $ \kappa $ , together with the ratio  of
writhe to twist $ \chi_W / {T}_{w} $ as functions
of the scaled supercoiling variable $ \eta=  {\chi \, A \over L} $. These theoretical curves
have been obtained with the same parameters as the "hat" curves of Figure \ref{fig3}. They
show a very rapid change of behaviour, 
a quasi-transition,  for $ {\eta}_c \approx 1.0$, with the following
two very different regimes:
\begin{itemize}
\item 
   Below ${\eta}_c $
the twist of the DNA chain increases linearly with the supercoiling $\eta$,
as in  a non flexible  rod with  an effective twist rigidity 
$ C_{eff}=.82\, C$. The ratio of writhe to twist stays almost constant at the value $0.2$.
\item
  Above ${\eta}_c $ the torque depends  weakly on the
  supercoiling $\eta$, the twist becomes nearly constant while 
   the writhe increases linearly with the supercoiling.
\end{itemize}
The behaviour in the large supercoiling regime is 
 reminiscent of the mechanical instability leading to
the formation of plectonemes which is easily observed by manipulating  macroscopic
elastic rods such as telephone cords. One must be careful with this analogy because the
plectonemic instability corresponds to a zero temperature limit,
 while in the case of DNA, much of the elasticity comes from entropic effects
 and  thermal fluctuations play a crucial role. 
 Yet we would like to point towards the existence, in the large 
 supercoiling regime of the RLC, of  excitations which share some of the properties
 of plectonemes.
 
We define  $ {\cal E}_p $  as a set of  undeformed plectonemes having the axis of
 their winding {\it   helices   arbitrarily oriented with respect to the force direction }
( $z-$ axis ).
  We then  introduce the angular  distribution $ P( \theta ) $  of the tangent vector $\vt$ 
 about  $z-$ axis,  averaged along the plectoneme  and upon the    
   set $ {\cal E}_p $. Let us sketch  the proof of the symmetry relation:
 $  P( \theta ) = P(\pi - \theta ) $. 
Introducing the  angular distribution $ {\cal P} ( {\alpha}_p ) $ of the plectonemes axis
and the tangent vector  $ {\vt }_{plect}(s) $ relative to  a vertical plectoneme,
 $ P( \theta )$  reads  as follows :
$$ 
 P( \theta )= \int_0^{\pi}  {\cal P} ( {\alpha}_p) d {\alpha}_p {1 \over L}
\int_0^{L}  ds \,\delta \(( \cos  \theta - \vz \cdot  R\(( \vy,{\alpha}_p  \)) {\vt }_{plect}(s)\))
$$
Ignoring the  plectoneme handle contribution, we  can  rewrite $ P( \theta ) $ in terms of 
the vertical  solenoid  $\phi$-dependent tangent vector
  $ {\vt }_{sol}(\phi,{\theta}_0)=  R( \vz,\phi ) R( \vy,{\theta}_0 )\vz $ :
$$ 
 P( \theta )= \int_0^{\pi}  {\cal P} ( {\alpha}_p) d {\alpha}_p  {1 \over 4 n_0 \pi}
\int_0^{ 2 n_0 \pi } d \phi \((  \delta ( \cos  \theta - 
\vu ({\alpha}_p)\cdot {\vt }_{sol}(\phi,{\theta}_0) ) + (\phi \rightarrow \pi-\phi, 
 {\theta}_0 \rightarrow \pi - {\theta}_0 ) \))
$$
where  we have used  the idendity $ \vA\cdot R \vB= R^{-1}\vA \cdot \vB $ and 
defined the unit vector $ \vu ({\alpha}_p) =  R( \vy,-{\alpha}_p ) \vz $.  
Then we compute: 
$\vu ({\alpha}_p)\cdot {\vt }_{sol}(\phi,{\theta}_0)=
 \sin{\alpha}_p \cos \phi \sin {\theta}_0 + \cos {\alpha}_p  \cos {\theta}_0 $ and 
  by simple inspection, we arrive to the desired relation. Because  of the 
$ \phi$  averaging,  the proof holds true if one takes an arbitrary axis in the $ z= 0$
plane instead of the $ y $ axis to rotate the plectoneme.  
In practice the plectonemes  are deformed by the thermal Brownian motion. More complex 
structures can also appear, like branched plectonemes. It looks however 
reasonable to assume that the above symmetry property is not affected, on average, 
 by  thermal fluctuations. 

In order to measure the degree of symmetry of the distribution of
$\theta$ angle along the chain, we
 introduce the function $ plecto( \eta) $ defined as follows:
\be
   plecto( \eta)= {\int_0  ^{\pi}   P( \theta ,\eta)   P( \pi- \theta ,\eta) d(\cos\theta) \over
\int_0  ^{\pi}   { P( \theta ,\eta) }^2 d(\cos\theta) }
\label{plecto}
\ee
It is clear that $ plecto( \eta) $   reduces to unity  
for pure plectonemic configurations  and goes to zero
 in the limit of   rectilinear  chains.
Within the RLC model, the probablilty  $P( \theta,\eta ) $ 
 is given by the thermal average  
of the  molecular axis angular distribution  when one runs  along the molecular chain. 
Exploiting the quantum mechanics  analogy,  
it is easily proved   that $ P( \theta,\eta ) $ 
 is equal to $ {\Psi}_0(\theta)^2$, the square of the
ground state wave function  introduced in subsection V.B.  
This has been used to compute the function {\it plecto}$(\eta)$ which
is plotted in figure (\ref{figplecto}).
The shape of $ P( \theta ,\eta ) $ changes in a very characteristic way 
when ones goes from  $ \eta=0 $ to $\eta \approx 4$. 
When  $ 0 \leq \, \eta  \leq {\eta}_c  \approx 1$,  $ P( \theta ,\eta ) $ has
a rather  narrow peak at $ \theta=0 $ with  a nearly  vanishing tail for 
  $  \theta > {\pi \over 2} $. As a consequence, the function $ plecto(\eta )$ 
is practically null  within  this interval.
 A secondary peak at  $ \theta= \pi $ begins to develop 
when $  \eta  \geq {\eta}_c $ and it  reaches  about the same height as the primary peak 
at ${\eta}  \approx 4 $. The building of this two bumps structure  is accompanied by 
an almost linear increase of the  function $ plecto(\eta )$ which reaches the value $ 0.9 $
near $ \eta= 4 $. This behaviour  suggests that
 thermally deformed plectoneme-like configurations
are responsible for the sharp increase of the  writhe-to-twist ratio and the flattening 
of the  curve torque versus supercoiling  above the critical value ${\eta}_c \approx 1$.
 Finally a useful piece of  information about the quasi-transition  near
$  \eta  \geq {\eta}_c $ is the study of  the variation of the energy gap between
the groundstate and the first excited  state $ \Delta\eps= {\eps}_1 - {\eps}_0$, obtained 
by the method given in subsection V.B.2. 
The corresponding curve appears in Figure \ref{figplecto}
under the label "Gap". $ \Delta \eps $ is a decreasing function  of $ \eta$,
with a rather  sudden   fall  near  $ \eta \approx{\eta}_c $. 
This somewhat technical feature  has  a nice physical interpretation in terms of
the correlation length associated with the $ \cos\theta $ fluctuations,
 {\it at fixed torque }, which is given  by $ A/\Delta \eps$.
The jump of this correlation length  around $\eta_c$,
 is a further sign of  a fast change of physical regime.

The two regimes of supercoiling displayed on Figure \ref{figplecto}
  are   related, within the  quantum mechanical formalism of sect. V.B.1, to the  
double well structure  of the potential
$  V_{r} ( \alpha,k^2,\theta )$ when $ k^2 =-{\kappa}^2$.
On one hand, the potential has a relatively 
shallow well at $\theta =0 $, associated with the 
stretching potential energy. On the other hand, 
the regularized "writhe" potential produces
a deep narrow hole at  $\theta =\pi $.
Let us  call $ {\eps}_a,  \Psi_a  $  and $ {\eps}_b,  \Psi_b $ the eigenvalues and 
 eigenstates  associated  respectively  with the  semi-classical states localized
 in the two potential wells. As  $\kappa $  is increasing 
the energy levels  $ {\eps}_a$ and $ {\eps}_b $  are approaching each other.
 The level crossing  is avoided  near $\kappa_c $  
by  a quantum tunneling through the finite height
wall between the two  wells. This accounts for the peculiar variation of $ \Delta \eps $
 near $\eta_c$. The mixing of the two levels 
$ \Psi_a  $  and $ \Psi_b  $ by quantum tunnelling
generates the secondary bump in the probability distribution $ P(\theta )$.
 In the vicinity of the near crossing point, the groundstate energy ${\eps}_{0}$   
depends almost linearly upon the tunneling amplitude  $ t_{ab}$, 
  which  has typically a very rapid variation with $ \kappa^2 $. 
Since the ratio $ \eta / \kappa  $ is a linear function
 of ${\partial \eps_0 \over \partial\kappa^2}$  
( see  equation (\ref{etakappa}) ), $ \eta $ should  exhibit  a very sharp increase  
with $\kappa \ge \kappa_c \approx 1.4$ as is easily seen
 by turning   Figure \ref{figplecto}  by 90 degrees. 


\begin{center}
\section{ Fluctuations of the Extension in  a Supercoiled  DNA  Molecule}
\label{fluctuations}
 \end{center}
 
\begin{figure}
\centerline{\epsfxsize=8cm
\epsfysize=14cm
\epsffile{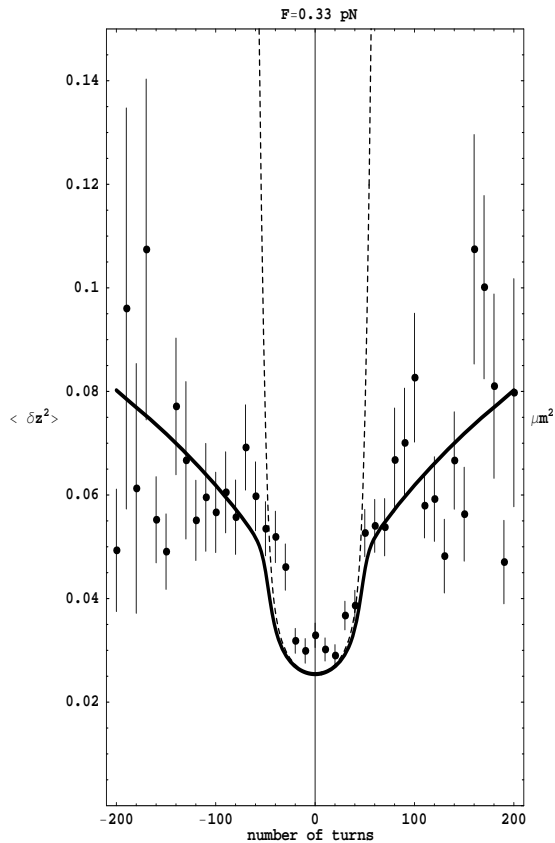}
 }
\caption{Mean square fluctuations of DNA  supercoiled  molecule extension
 for $F=0.33 \, pN$. The points are the experimental data.
 The thick continuous line is the theory, for the experimental
 situation where 
 the supercoiling turn number $ n = {L \eta \over 2 \pi A } \approx 50 \,\eta $ is fixed. 
 The dotted line is the theoretical result for another situation
 where the torque would be fixed  (the number of turns  is then to be
 understood as  a thermal average value).   }   
\label{figfluct}
\end{figure}

In this section we would like to give  a brief  analysis of preliminary measurements
 of  the supercoiled 
DNA extension fluctuations, performed at a force $ F=0.33 \, pN $ by the experimental group
in the ENS.
 We shall compare   the experimental results  with   the $ RLC $  
model predictions using the  parameters $ b/A $ and $ C/A $ deduced  from 
the extension versus supercoiling curves analysis, performed in 
section \ref{ana_exp}.
The  mean square deviations of the fluctuations 
of the molecule extension along the force direction is  given by
the second derivative of the free energy $ {\cal F}= -k_B T \log Z(F) $
 with respect to F:
\begin{equation} 
 \langle \, \delta\,z ^2 \, \rangle = \langle \,z^2  \, \rangle-
 {\langle \,z  \, \rangle }^2 = (k_B \, T)^2 { {\partial}^2 \log Z \over \partial F^2 }
\end{equation} 
It is convenient to replace  the first derivative $ { \partial \log Z \over \partial F }$
by its expression in terms of 
the average relative extension using equation  (\ref{extens}). The  extension mean square
fluctuation can then be written as follows:
\be 
 \langle \, \delta\,z ^2 \, \rangle =
 L \(( { k_B T \over F}\))  \, \alpha \, {  \partial \over \partial \alpha } 
 \(( {  \langle \,z  \, \rangle  \over L }  \)) 
 \label{fluctzkappa}                                  
 \ee
 We have verified that the value of $ \langle \, \delta\,z ^2 \, \rangle $ obtained in
 the discretized  RLC model in the limit of no supercoiling, $\chi=0$,
 agrees  with the one given by the WLC model  to better than $ 3 \, 10^{-3}  $.
  In the situation where the free end of the molecular chain is subjected to a supercoiling
 constraint  the fluctuations will be different depending
 on whether  the measurement
 is performed  at fixed torque $\kappa$ or at fixed supercoiling  angle $\chi$.
 In the first case the result is readily obtained  from equation (\ref{extens}) :
 \be
 \langle \, \delta\,z ^2 \, \rangle {\vert}_{\kappa} = - L \(( { k_B T \over F}\)) 
 \,{ \alpha \, {\partial}^2 {\epsilon}_0(\alpha,-\kappa^2)\over \partial {\alpha }^2 } 
 \ee 
 In the actual experiment   the extension  fluctuations are  
measured   at fixed $ \chi $ and an extra term has to be added to the above expression
due to the fact that the torque is now  a function of $ \alpha $ and $ \eta $:
 \be
\langle \, \delta\,z ^2 \, \rangle {\vert}_{\eta} =  L \(( { k_B T \over F}\))\, \alpha \,\((
-  { {\partial}^2 {\epsilon}_0(\alpha,-\kappa^2)\over \partial {\alpha }^2 } + 
  2 \kappa {\partial \kappa \over \partial \alpha  }
 { {\partial}^2 {\epsilon}_0(\alpha,-\kappa^2)\over \partial \alpha \partial k^2 } \))
  \ee
Using equation (\ref{etakappa}), the  extra term  can be transformed by writing:\\
$ 2 {\partial \eps_0 \over \partial k^2}(\alpha,-\kappa^2)={\eta
\over \kappa} -A/C $. 
 One gets finally   the  mean square extension fluctuations  at fixed supercoiling angle:
\be 
  \langle \, \delta\,z ^2 \, \rangle{\vert}_{\eta} =
 \langle \, \delta\,z ^2 \, \rangle {\vert}_{\kappa}- L \(( { k_B T \over F }\))
\, \alpha \,{\eta \over \kappa } { \(( {  \partial \kappa \over \partial \alpha  } \)) }^2 
\label{fluctzeta}
\ee
In the above formula the second term is clearly negative so we expect that
 $ \langle \, \delta\,z ^2 \, \rangle{\vert}_{\eta} <
 \langle \, \delta\,z ^2 \, \rangle {\vert}_{\kappa}$. The curves giving  
$ \langle \, \delta\,z ^2 \, \rangle $ 
versus the number of supercoils  $ n = {L \eta \over 2 \pi A } \approx 50 \,\eta $
are displayed on Figure \ref{figfluct},  $ \langle \, \delta\,z ^2 \, \rangle{\vert}_{\eta} $
as a thick continuous line and  $ \langle \, \delta\,z ^2 \, \rangle{\vert}_{\kappa} $ as 
 a dashed line. It appears clearly that 
 $  \langle \, \delta\,z ^2 \, \rangle{\vert}_{\eta} \ll 
 \langle \, \delta\,z ^2 \, \rangle {\vert}_{\kappa} $, when $ |\eta |> {\eta}_c \approx 1 $.  
This implies  a strong  cancellation 
between the two terms of  formula  (\ref{fluctzeta}), which
is then not suitable  for an evaluation of 
 $  \langle \, \delta\,z ^2 \, \rangle{\vert}_{\eta} $. 
  To get the thick solid curve of Figure \ref{figfluct}, we have used the fact 
 that the  calculation procedure  giving, at  reduced force $ \alpha$,
  the relative  extension   versus  the supercoiling angle, 
 $ {  \langle \,z  \, \rangle  \over L } (\alpha ,\eta) $, is precise enough to 
allow the evaluation of the  partial derivative with respect to $\alpha$ by
 a three points  finite difference formula.

 The difference between the two statistical ensembles 
  can be understood qualitatively  from the previous section considerations
 and specially by looking  at the curves 
 of  Figure \ref{figplecto}. In a situation where the torque is fixed  
at $ \kappa = {\kappa}_c \approx 1.4 $ the supercoiling angle $ \chi $
 is practically unconstrained. It can  fluctuate rather freely through
 thermal excitation of plectonemes, which have no effect upon the torque.
 In contrast   when $ \eta $ is fixed  at a value $  \eta > {\eta}_c \approx 1 $ 
the ensemble is more constrained. The torque $ \kappa $ is still practically fixed 
 at the critical  value $ {\kappa}_c  $. As a consequence,  
the    writhe $ \chi_W= \chi - T_w \approx \chi - {\kappa}_c L/C  $ has 
 approximately  a fixed value  and  then  the only plectonemes which
 can be created  are  those   having a very limited  
  range of writhe. 

 As  is apparent on Figure \ref{figfluct}, the experimental points
 agree reasonably well with  the RLC model predictions,
 taking into account  the  quoted uncertainties, which are
exclusively  of statistical origin. ( Systematic effects
are smaller). 
We notice the quasi-critical jump  of 
$ {  \langle \,z  \, \rangle  \over L } (\alpha ,\eta) $,
 near $ n_c \approx 50 $, which is 
also present in  the data.

\section{Summary and Conclusion}
\label{conclusion}
Together with new relevant contributions, this paper gives a detailed account of a 
  work  of the authors, which appeared 
 previously under a very concise letter form \cite{BM1}. It was
a rather successful  attempt to describe the entropic elasticity  of supercoiled 
single DNA molecule in terms of the thermal fluctuations of an elastic rod. 

As in most works based upon  a similar model, the rigidity tensor was assumed 
to be symmetric under rotations about the molecular axis and then fully
described by two elastic constants, the bending  and twist rigidities.  
Because of the DNA  helical structure, one may have expected an axially asymmetric
rigidity tensor, involving extra elastic constants. We have proved  in this paper that 
axial symmetry breaking  contributions to the elastic energy  
are averaged out upon a " coarse graining " involving a  length resolution $ \Delta l $ about
 three  times the double helix pitch $ p $. Such a  value corresponds  
to the actual experimental resolution.
The  model developed in the present paper is
not expected to be realistic  at length scale
 below twice the DNA double helix pitch.

To implement in the RLC  partition function the supercoiling empirical constraint $\chi$,
we have followed a direct procedure, instead of  adapting to open molecular DNA chains
the formalism used in the study of supercoiled plasmids.
 Imposing a well defined regularity condition
upon the Euler angles describing  locally the molecular chain, we have been able to 
decompose the  empirical supercoiling angle $\chi $ as a sum of two line integrals  
 $ T_w $ and  $\chi_W $, associated
 respectively with the twist and writhe  contributions.
The  local writhe  $\chi_W $ is a  line integral taken 
 along  regular trajectories   of the  chain unit tangent vector $\vt$. They are drawn
upon the  unit sphere $ S_2 $ pierced by a hole having a small radius $\epsilon$,
 localized at the south pole defined by the direction of the
stretching force direction. Such a prescription forbids the crossing 
 of a spherical coordinate singularity   which
   would invalidate our derivation of  the local  writhe  formula. 
 When it is applied in compliance with the above prescription, 
 the  local  writhe formula allows correct evaluations  of the   writhe of solenoid and plectoneme
 configurations, having  an  arbitrarily space orientation. More generally
 the local writhe  formula, though not explicitly rotation invariant,  is
shown to   lead, in the small $\epsilon$ limit, to  rotation invariant results.

The  partition function  of the RLC having a given  supercoiling angle $\chi $  is written 
as  the  convolution product of the partition function $ Z_T( T_w) $ of a twisted non flexible rod
times the partition function  $ Z_W( \chi_W ) $  of a worm like chain having a fixed  writhe angle.
 The Fourier transform $ \tilde Z_W(k) $ is shown to be the analytic continuation to imaginary
time of the Feynman amplitude  describing  the quantum evolution of unit charge
particle moving upon a sphere upon the joint action of an electric dc field and
a magnetic monopole having  an unquantized charge $k$. The  associated Hamiltonian 
$ {\hat H}_{RLC}(k) $ is readily obtained from standard Quantum Mechanics rules.

As in the  quantum magnetic monopole problem, we have found a pathology which
 is associated 
 with  the singular
behaviour of the $  {\hat H}_{RLC}(k) $ potential term near the south pole.
 We have  proved that   an angular cutoff  is generated
by   a discretization of the chain involving
 an elementary link $b$ ( this result was given
 without proof in our previous work  \cite{BM1}).
We have derived  analytically from the discretized elastic energy 
a regularized  version of RLC continuous model. The singular writhe potential 
is  multiplied by a  regulating function going smoothly to zero 
when  $\sin^2\theta \leq {b\over A}$. We  have arrived in this
way  to a well behaved  Hamiltonian  $ {\hat H}_{RLC}^{r}(k)$, which provided a precise  
mathematical definition of RLC model used in this paper.  It  leads to a  partition function 
free of any pathology. In an independent work following ours\cite{BM1},
  Moroz and  Nelson proposed a high force  perturbation  method \cite{mornel1}
 where no angular cutoff  is
introduced from the start. To finite order, the divergent perturbation series, 
hopefully asymptotic,  does not 
 see the singularity at the south pole, though it is always present in the model.
 In order to make contact  with experiment, the authors have to impose 
upon their expansion parameter a sharp " technical " bound. 
 Its effect is  to restrict considerably the
domain  of force and supercoiling where a comparison with experiment is possible.
 This limits the precision of their determination of $C$.

 Within our  regularized RLC model, we have developed methods  
allowing the  computation,  to an arbitrary precision, of  the relative extension 
versus supercoiling curves, the so called " hat " curves. 
The Fourier transform involved in the partition function
is evaluated by the saddle point method,  in the limit of a contour length $L$
much larger than  the persistence  length $A$. It leads to a parametric 
representation of the hat curves in terms of the torque acting upon the molecule's 
free end.  The final theoretical ingredient is the ground state energy  of the regulated RLC 
Hamiltonian. It is obtained following  two  methods:  {\it a) } an explicit solution of the 
Schr\"odinger equation associated with $ {\hat H}_{RLC}^{r}(k) $,  {\it b) } the iteration of
 the transfer matrix deduced directly from the discretized  elastic energy. Though they 
are not strictly equivalent from a mathematical point, 
they lead  to identical results to few $\%$.
 
 Our data analysis involves three values of the stretching forces:

 $F=0.116, \,0.197, \,0.328\,pN$.
For each hat curve point and a fixed  ${b\over A}$, the RLC model leads to an empirical
 value of the ratio $ {C \over A}$  as  function of
the measured relative extension and supercoiling angle. 
 If the model  provides an adequate description of  the hat curve,
 the set of  empirical values should cluster around the actual value $ {C \over A}$.  
For each cluster we have plotted  the  mean value   $ \langle {C \over A} \rangle$  
 and the variance  $ \sigma_r $  versus   $ {b\over A}$.  The  best value $ {b \over A}=0.14$ 
   corresponds  to the minimum  of  the ratio $ \sigma_r/\langle {C \over A} \rangle $, which 
measures the ability of the model to fit the data.  The preferred cutoff length $b$, 
is found to be  about two times the 
double helix pitch and it is close to the length resolution 
$ \Delta l $ invoked to suppress  axially  asymmetric 
elastic energy terms. Our determination of ${C \over A}= 1.64 \pm 0.04 $  is obtained 
  from a weighted average of $ \langle {C \over A} \rangle $ relative to 
 the three forces involved in the fit, taking  $ {b \over A}=0.14$. 
 This number turns out to be  remarkably stable under  variations of $b$  
within the range $ 0.08 A \leq  b \leq  0.2  A$: the  
deviations of  ${C \over A}$ from 1.64  stay   below the level of $ 8 \%  $. 
In contrast, the quality of the  hat curves fit 
which  is very satisfactory at  the   best $b$ value 
becomes poorer away from it, specially for low $b$ values.
 Our central value  for the twist rigidity $ C=84 \pm  10 \,nm$ 
differs by about $ 25 \% $ from the ones 
given by Moroz and Nelson \cite{mornel2}. The absence of any uncertainty estimate 
 and an analyzed  data set having a small overlap with ours makes the
physical significance of this apparent discrepancy   difficult  to assess. 

Although these quantities  are not yet directly accessible to experiment, 
we have computed, as functions of the supercoiling  $\chi $, 
  the torque $ \Gamma $  and the  average twist
 $ \langle T_w \rangle$ (or equivalently
the  average writhe $ \langle \chi_W \rangle= \chi-  \langle T_w \rangle$). Use 
has been made of a remarkable property:  $ \langle T_w \rangle$ is related to  $ \Gamma $
by the linear elasticity formula for a  non-flexible twisted rod. 

In  the case of the highest force ($ F= 0.33 \,pN $),  the  two  curves,
 torque and writhe to twist ratio  versus supercoiling,  exhibit a rather sharp change 
 of regime near the same critical value $ \chi_c $ of the supercoiling angle:
 the torque, after a nearly linear increase, becomes almost 
supercoiling independent while the  writhe to twist ratio,
 initialy confined to the $ 20 \% $ level develops a fast linear increase.
 This behaviour is reminiscent of the buckling instability  of a twisted rubber tube 
associated with the creation of plectonemes, able to absorb supercoiling at constant torque.
We have shown that the configurations excited in the RLC model
 above the critical value $ \chi_c $  share a 
simple global symmetry  property with a set of undeformed plectonemes    
arbitrarily oriented  with respect to the  force direction $z$ axis: 
 the angular distribution  of the tangent vector $ \vt $ has a forward backward symmetry 
with respect to the $z$ axis.  The existence of this rather sharp cross over  
has been confirmed by an analysis of the  extension fluctuations versus  supercoiling:
the predicted fluctuations jump near  $ \chi_c $
   is clearly seen on the preliminary experimental data.
 
The overall good  agreement of our predictions with
 the analyzed experimental data seem to 
indicate that the self-avoiding effects, not included in our RLC model,  play a limited role in 
the low supercoiling regime $ \vert \sigma \vert  \leq 0.02 $. This was suggested by the Monte
Carlo  simulations of Marko and Vologodskii \cite{marvol} who use
two infinite  impenetrable walls  to
adapt the closed chain formalism  to the supercoiled open chains.
A further test follows from  computing $ \lim_{F=0}\,  \langle \,\chi_{W} \, \rangle / \chi $
for $\sigma $ values taken in the range: $ 0 \le -\sigma \le 0.04 $. 
This ratio   compares well to the closed DNA chain  
 writhe  average $ < Wr/\Delta Lk > $  
obtained  by Monte Carlo simulations \cite{volog}. 
The two results diverge by less than $ 10\% $ when $ \vert \sigma \vert \leq 0.04 $
and both agree with the experimental data within errors \cite{boles},\cite{adrian}. 
It should be said that the two calculations use different formulas for the writhe:
an open chain  local version  in  the present paper, 
the non-local loop  geometrical formula \cite{geometry} in reference \cite{volog} .
 
    In view of the wealth of experimental  information, 
there are strong motivations   to extend the  validity
of the present RLC model to  a larger domain
of the $( F,\sigma)$ plane. Further work has to be pursued in
several directions. Non-local constraints in the chain tangent vector $\vt $ space  have to 
be implemented. There are first the empirical geometrical constraints associated
with the DNA anchoring glass plate and  the finite radius of the tracking bead.
The self-avoiding effects induced  by the Coulomb  repulsion within  the DNA 
chain have to be studied  for  the range of ionic strength accessible to  experiments.
The only practical  approach to these problems seem, for the moment, the 
Monte Carlo simulations technique.

 Another interesting perspective is  to incorporate in the present model
 the double helix  structure  the DNA, in order to describe the
DNA  denaturation transition induced  by negative
supercoling above $ F=0.5 \,pN$. Some
first steps in this direction have already been taken in \cite{liverpool,zhou}.
Recently a model coupling the hydrogen-bond opening with the untwisting of the 
double helix has been proposed \cite{cocco}. It allows a unified
description of DNA denaturation driven by thermal fluctuations or induced by 
the double helix untwisting, in the case a straight line molecular chain. 
It will be  of interest to combine this model  with the elastic  RLC model 
in order to  study the onset of the  denaturation  transition
 at moderate stretching forces, say below $ 1 pN$,
 where bending fluctuations can no longer  be neglected.

\section*{Acknowledgments}
It is a great pleasure to thank all members of the experimental group
at ENS, J.-F. Allemand, D. Bensimon, V. Croquette and T.R. Strick,
for many valuable discussion, and for providing us some of their  
unpublished data. We have  benefited from discussions
with J.-P. Bouchaud, A. Comtet and C. Monthus on the winding
of random walks. We are very grateful to D. Bensimon
and M.A. Bouchiat for their careful reading of 
this long  manuscript.  The work of MM was supported in part by the
National Science Foundation under grant No. PHY94-07194.
\begin{appendix}
\section{  Rotation Matrix Algebra}
Let us denote by $ R( \vn, \gamma ) $ the rotation of angle $\gamma$ about the 
unitary
vector $ \vn $. With this notation, the rotation $ {\cal R}(s) $, which 
specifies  
an arbitrary  DNA chain configuration in terms of the three  Euler angles, is 
given by :
\be 
 {\cal R}(s) = R\(( \vz, \phi(s) \))  R\(( \vy, \theta(s) \)) R\(( \vz, \psi(s) 
\)) 
\label{Re}
 \ee
 Another very useful way of writing $ {\cal R}(s) $  follows from  
 the rotation group relation:
\be 
\hat{R}=R_1^{-1} R( \vn, \gamma ) R_1= R( R_1^{-1} \vn,\gamma )
\label{grR}
\ee
The  proof of (\ref{grR}) follows from  basic properties of a rotation matrix:
first, the rotation axis is   the rotation matrix eigenvector with unit eigenvalue; 
we  verify that it is indeed the case for $ R_1^{-1} \vn$:
 $\hat{R} R_1^{-1} \vn= R_1^{-1} R( \vn, \gamma ) \vn= R_1^{-1} \vn $;
second,  the rotations $\hat{R}$ and R$( \vn, \gamma )$ have, by construction, 
the same
eigenvalues: $ 1, \exp (i \gamma ),  \exp (-i \gamma )$ 
and hence the same rotation angle $\gamma$. The new  form of $ {\cal R}(s) $
is then obtained  from the simple manipulations:
\bea 
  {\cal R}(s) &= & R\(( \vz, \phi \))  R\(( \vy, \theta \)) R\(( \vz, \psi \)) 
\nonumber \\
&=& R\(( \vz, \phi +\psi \))  R^{-1}\(( \vz, \psi \)) R\(( \vy, \theta \))
 R\(( \vz, \psi \)) \nonumber \\
&=& R\(( \vz, \phi +\psi \))  R\(( R( \vz, -\psi ) \vy, \theta \)) 
\eea
We are going now to discuss the two angular velocity vectors
 and $ \Om (s) $ and $ \Up (s)$ defined by  relations valid 
for arbitrary vectors  $ \vX $ and $ \vY $:
\bea
     \dot {\cal R} (s)   {\cal R}^{-1}(s)  \vX  &=&    \Om (s)  \wedge \vX   
   \label{Om}\\
   {\cal R}^{-1}(s)  \dot {\cal R} (s)    \vY  &=&    \Up(s)  \wedge \vY  
    \label{Up}
\eea
 Applying  to equation (\ref{Up}) the simple identity 
$ {\cal R}\((\va\wedge\vb\))=\(({\cal R}\va\))\wedge \(({\cal R}\vb\)) $
and taking $ \vY=  {\cal R}^{-1}(s) \vX $,  one gets immediately the relation:
$\Om (s)  =  {\cal R} (s)  \Up(s)$. It facilitates the evaluation of the 
components of 
$ \Om (s)  $ upon the moving trihedron $\{ {\ve}_{i}(s)\} $ since we can write:
\be  
 {\Omega}_{i}= \Om(s) \cdot {\ve}_{i}(s) =\Up(s) \cdot {\ve}_{i}^0(s)
\label{omi} 
\ee
To simplify the writing in the explicit computation of $ \Up(s) $, we 
introduce the notations :
$ R_1=  R\(( \vz, \phi(s) \)) ,  R_2=  R\(( \vy, \theta(s) \)) ,  R_3=R\(( \vz, 
\psi(s) \))$.
With some elementary matrix algebra we get:
$$  {\cal R}^{-1}(s)  \dot {\cal R} (s)  = ( R_2 R_3)^{-1} R_1^{-1} \dot R_1 R_2 
R_3+
R_3^{-1} R_2^{-1} \dot R_2 R_3+ R_3 \dot R_3 $$
As an intermediary step,  we compute :
$$ R_a^{-1} R^{-1}(\vn,\gamma(s) ) \dot R (\vn,\gamma(s) )  R_a \vX= 
 R_a^{-1} \(( \dot \gamma(s) \vn \wedge  R_a \vX \))= \dot \gamma  R_a^{-1} \vn 
\wedge \vX $$
This result is nothing but the relation (\ref{grR})
applied to an infinitesimal rotation.
     Using the above results we arrive finally to the following expression for $ 
\Up (s) $
\be 
 \Up (s)= \dot \phi  R\(( \vz, -\psi \))  R\(( \vy, -\theta \))  \vz +            
                       \dot \theta R\(( \vz, -\psi \)) \vy + \dot \psi   \vz
\ee
It is now  convenient to decompose $ \Up (s) $ in a longitudinal  
${\Up}_{\parallel} (s)$
and a transverse part  ${\Up}_{\perp} (s)$
\bea
{\Up}_{\parallel} &=&( \cos\theta \, \dot \phi + \dot \psi  ) \vz  \\
{\Up}_{\perp} &=&  R \(( \vz, -\psi \)) ( -\dot \phi \, \sin\theta\vx + \dot 
\theta\,  \vy ) 
\eea
 We get immediately the quantities appearing in the RLC elastic energy:
\bea
 {\Omega}_3 &=& \cos\theta \, \dot \phi + \dot \psi \\
 {\Omega_{\perp}}^2 &=& {\Up}_{\perp}^2= {\Omega}_1^2+{\Omega}_2^2 =
 \dot \phi^2 \, \sin^2\theta + \dot \theta^2
\eea 
Our next step is to compute the  cylindrical  symmetry breaking term 
$ \Delta \Omega(s)=  {\Omega}_{1}^2-{\Omega}_{2}^2 $.
 Introducing the angle 
 $\zeta(s) =\arctan\((\dot{\theta}/(\sin\theta\dot{\phi} )\))$,
we can write: 
$ {\Up}_{\perp} = -\Omega_{\perp} \,R( \vz,-\psi-\zeta)\,\vx $.
A physical interpretation of  $\zeta(s)$ is obtained by computing the $s$ 
derivative of 
the tangent unit vector $ \vt(s) $:
\bea
{d \vt(s) \over d s} &= & \Om \wedge \vt = {\cal R}(s)
 \(( {\Up}_{\perp} \wedge \vz \)) \nonumber \\
&=& -\Omega_{\perp}  {\cal R}(s) \,R( \vz,-\psi-\zeta) \((\vx  \wedge \vz \))
 \nonumber \\
&= & \Omega_{\perp} R\(( \vz, \phi \))  R\(( \vy, \theta \)) R\(( \vz, -\zeta \)) \vy
\nonumber
\eea
We see that $-\zeta(s)$ plays  the role of the Euler 
 angle $\psi(s) $  vis-\`a-vis the Seret-Frenet trihedron so  that $\zeta(s)$ is 
clearly
connected with the writhe.  
It is now a simple matter to get the component  ${\Omega}_1 $
  by writing the series of equalities:
\bea
{\Omega}_1 &=& \Up(s) \cdot {\ve}_{1}^0(s)=
 -\Omega_{\perp} \(( R( \vz,-\psi-\zeta) \,\vx \)) \cdot \(( R( \vz,\omega_0 s)\,\vx \))
 \nonumber \\
&=&
-\Omega_{\perp} \vx \cdot \(( R( \vz,\psi+\zeta +\omega_0 s)\,\vx \))=
 -\Omega_{\perp}\,\cos(\psi+\zeta +\omega_0 s) 
\nonumber
\eea
In a  similar way we obtain ${\Omega}_1 = -\Omega_{\perp}\,\sin(\psi+\zeta 
+\omega_0 s)$
and we arrive finally at the following expression for $ \Delta \Omega(s)$:
\be
\Delta \Omega(s)=  {\Omega}_{1}^2-{\Omega}_{2}^2 = 
 {\Omega_{\perp}}^2 \cos 2 (\psi+\zeta +\omega_0 s)
\ee

Introducing  the  length resolution function 
$ P(s)={1\over \sqrt{2 \pi } \ell}\exp( -{1\over2} s^2 
/{\ell}^2) $ ( Note the change of notation: 
 $\ell$ stands for $\Delta\,l$ used in the main body of the paper), we proceed  with 
the computation of 
the average  $ \overline{  \Delta\Omega(s) }$:
$$
 \overline{  \Delta\Omega(s) }=\int ds_1  P(s_1-s)\Delta\Omega(s_1) 
      = {1\over \sqrt{2 \pi } } \int du \exp \((-\frac{1}{2} u^2\)) 
       \Delta\Omega(s + u \ell)  $$

Let us first  neglect the variation of  $\Omega_{\perp} $ within the interval
$( s-\ell,s +\ell )$  and perform a first order expansion
 in $\ell $ of the phase  $ \psi(s+ u \ell)+\zeta (s+ u \ell) $;
 this is justified  since 
its  variation   is expected to be of the order of $ { \ell\over A } \approx 0.2 
 $ 
( We have  proved  explicitly  the thermal average inequality: 
$ \langle \dot \psi \rangle / \eta   <  1 /A $).
 In this way $\overline{\Delta\Omega(s) }$
is transformed into a Gauss integral:
\bea
   \overline{  \Delta\Omega(s) } &= & \Delta \Omega(s) {1\over \sqrt{2 \pi } }
\int du \exp \((-\frac{1}{2} u^2 \))
\cos\(( 2 u \ell( \dot \psi+ \dot \zeta +\omega_0 )\))  \nonumber \\
& = & \Delta \Omega(s) 
\exp\((  -2 \, {\ell}^2  \,(\dot \psi+ \dot \zeta +\omega_0 )^2 \)) 
\eea
Using the estimates $ \vert \dot \psi \vert / \omega_0   \vert \sim 
\vert \dot \zeta \vert / \omega_0  \sim \vert \chi \vert/( L \omega_0) 
=\vert\sigma\vert $  
and the fact that in the present paper our analysis is restricted to values of
$ \vert\sigma\vert < 4 \; 10^{-2} $,  we can write,
introducing the pitch $p=2\pi/\omega_0$ :  
\be 
\overline{  \Delta\Omega(s) }\approx \Delta\Omega(s) 
\exp\((  -\frac{1}{2} (\frac{4\pi \,\ell}{p})^2 \))
\ee 
Taking $ p=3.4 \, nm $, $\ell=10 nm $  we find: $\frac{4\pi \,\ell}{p} \approx 37$;
it means that $ \overline{  \Delta\Omega(s) }/ \Delta\Omega(s) $ is zero 
for all practical purposes. Using  instead $ \ell=b= 7 \, nm $
 would not make any difference.
 Let us say few words about the term $\delta \overline{  \Delta\Omega(s) }$
  involving the variation of  $ {\Omega_{\perp}}^2 $. A computation
similar to the previous one gives the following result:
$$ \delta \overline{  \Delta\Omega(s) } =
 { \partial \, {\Omega_{\perp}}^2  \over  \partial \,s}\,\ell
 \, \frac{4\pi \,\ell}{p}
\exp\((  -\frac{1}{2} (\frac{4\pi \,\ell}{p})^2 \)) 
\sin 2 (\psi+\zeta +\omega_0 s)$$
The rate of variation of $ {\Omega_{\perp}}^2 $, which is 
the inverse of the curvature radius square, is expected to be of the order
$ 1/A $ so that
 $  { 1 \over  {\Omega_{\perp}}^2 } {\partial \, {\Omega_{\perp}}^2  \over  
\partial \,s}
\,\ell \sim \ell /A =0.2 $. It follows that $\delta \overline{ \Delta\Omega(s)}$
is still exceedingly small compared to $\Delta\Omega(s)$. Futhermore it is 
easily shown
that  the thermal average derivative
$  { \partial \, \langle {\Omega_{\perp}}^2 \rangle  \over  \partial \,s}  $
 vanishes for $ s \gg A $.

\section{ The Symmetric Top and the RLC Model}

 In this appendix we shall use  
for convenience  a units system where 
  $ \hbar= c  =k_BT=1 $ and  write $ \Im{s}=t $. The elastic energy  $ E_{RLC} $ 
(eq. (2) and eq. (3) ) is transformed
by an   analytic continuation  towards the imaginary $s$ axis
 into $ -i $ times  the  action integral:
$$ 
{\cal A}(t_0,t_1) =   \int_{t_0}^{t_1} dt \(( {1\over2}\sum_{j=1}^3 {C}_j {\Omega}_{j}^2 
+f \cos \theta(t) \))
$$
where $ {C}_1={C}_2= A $, $ C_3=C$ and $ f= F/(k_BT) $. The  time 
derivatives  
 accounts for the relative change of sign between ${C}_i {\Omega}_{i}^2$ and 
the potentiel energy $ -f  \cos\theta(t) $. The analytically
 continued   partition 
function
 of the RLC model  is then identified with the Feynmann path integral amplitude:
$$ 
\langle {\theta}_1,{\phi}_1,{\psi}_1,t_1\vert{\theta}_0,{\phi}_0,{\psi}_0 ,t_0 
\rangle= 
\int {\cal D}\((  \theta,\phi,\psi \)) 
 \exp\((i  \int_{t_0}^{t_1} dt {\cal L}_{top} (t) \)) 
$$
The Lagrangian $ {\cal L}_{top} (t)=  {1\over2}\sum_1^3 {C}_i {\Omega}_{i}^2 +f \cos 
\theta(t) $
describes the motion of a spherical top with  inertia moments  $ {I}_i={C}_i $, 
under 
the action of a static electric field  $ E_0 $. (The  molecular electric moment 
is given by  $ f/E_0 $.) In order to compute explicitly the hamiltonian $ {\cal H}_{top} $
let us write $ {\cal L}_{top} (t)$ in terms of the Euler angles and their derivatives:
\be
 {\cal L}_{top}= {A \over 2} \left(
{\dot{\phi}}^2 \,{\sin^2\theta}+{ \dot{\theta}}^2 \right) +
 { C \over 2} \,( \dot{\psi} +\dot{\phi} \, \cos\theta )^2 +
f \cos \theta 
\ee 
One gets immediately the conjugate momentums relative to three Euler angles:
$ p_{\phi} = A \, \sin^2\theta  \,\dot{\phi} + 
\cos\theta \, p_{\psi}$,
$ p_{\psi}  =  C \,( \dot{\psi} +\dot{\phi} \, \cos\theta )$ and 
$p_{\theta} =  A \,\dot{\theta}$. The symmetric top hamiltonian is then 
readily obtained:
 \bea 
 {\cal H}_{top} &=&  {A \over 2} \left(
{\dot{\phi}}^2 \,{\sin^2\theta}+{ \dot{\theta}}^2 \right) +
 { C \over 2} \,( \dot{\psi} +\dot{\phi} \, \cos\theta )^2 -
f \cos \theta   \nonumber \\  
               &=& { (p_{\phi}-   \cos\theta \, p_{\psi} )^2 \over 2\,A\,\sin^2\theta }+
{ {p_{\theta}}^2 \over  2\,A}+{ {p_{\psi}}^2 \over  2\,C}-f \cos \theta
  \eea                                       
To get the hamiltonian operator $ {\hat{\cal H}}_{top} $ we apply the standard 
quantization rules:
$$
 p_{\phi} \rightarrow {\hat{p}}_{\phi} = { \partial \over i \partial \phi } \;,\;
      p_{\psi} \rightarrow {\hat{p}}_{\psi}={ \partial \over i \partial \psi } \;,\;
  p_{\theta}^2 \rightarrow {\hat{p}}_{\theta}^2=
-\frac{1}{ \sin\theta }\frac{\partial}{\partial\,\theta}\sin\theta \,
  \frac{\partial}{\partial \theta } 
$$
To get the partition function 
$ Z({\theta}_1,{\phi}_1,{\psi}_1,s_1\vert{\theta}_0,{\phi}_0,{\psi}_0,s_0 )$
 we expand the final and initial states upon eigenfuctions of the operators 
$ {\hat{p}}_{\phi}$ and ${\hat{p}}_{\psi}$: $ \exp(i\,m\,\phi +i\,k \, \psi ) $ 
where $ k $ and $ m $ are arbitrary real numbers, in contrast with the real
 symmmetric top
case, where they are integers. This difference, which follows from a detailed analysis
of the physics involved ( see section II B. for details), is responsible for the 
singular features  of the  continuous RLC model. The partition function with
the notations used in the paper to label the initial and final states  
reads as follows ;
\bea
   Z(\theta(L),\phi(L),\psi(L),L\vert \theta(0),0 ) & = &
\int\,dm \,dk  \exp\(( i\,m\,\phi(L) +i\,k \, \psi (L) \))  \nonumber \\  
& & \langle \theta(L)\vert \exp \(( -L\,\hat{\cal K} (k,m) \)) \vert \theta(0)\rangle 
\eea
 where the Hamiltonian $\hat{\cal K} (k,m) $ is given by:
$$ 
\hat{\cal K} (k,m)=
 -{1 \over  2\,A \,\sin\theta  } \, \frac{\partial}{\partial\,\theta}\sin\theta \,
 \frac{\partial}{\partial \theta }+
{ (m-  \cos\theta \, k )^{2} \over 2 \, A \,\sin^2 \theta }+{ k^2 \over  2\,C}-f \cos \theta
$$
The experimental  supercoiling constraint is implemented   by averaging
upon $ \phi(L) $ and  $ \psi (L) $  the above partition function 
 multiplied by  the Dirac function  $ \delta( \phi(L)+  \psi (L)-\chi ) $.
A straightforward computation gives:
$$ 
Z(\chi )= \int 
\,dk  \exp(i \,k\,\chi )
 \langle \theta(L)\vert \exp \(( -L\,\hat{\cal K} (k,k) \)) \vert  \theta(0)\rangle 
$$ 
The hamiltonian operator ${\hat H}_{RLC}(k)$ given by equation (16) 
is recovered by writing:
\bea 
{\hat H}_{RLC}(k) &=& {1\over A} \(( \hat{\cal K} (k,k)- {A\over 2\, C} k^2 \)) 
 \nonumber \\
&=&-\frac{1}{ 2\,\sin\theta }\frac{\partial}{\partial\,\theta}\sin\theta \,
  \frac{\partial}{\partial \theta }-
\alpha \cos\theta+{k^2 \over 2} {1-\cos\theta \over 1+\cos\theta}
 \eea 
The merit of this direct derivation is to show clearly that $ k$ is the unquantized angular
momentum  of the Euclidian  symmetric top problem  associated with
the RLC model. It is  the breaking of the Quantum Mechanics usual quantization rule
 which is at the origin  of the RLC model pathology.       
 \end{appendix}


\begin{references}
\vspace{-0.5cm}
\bibitem{smith}
S.B. Smith, L. Finzi and C. Bustamante, Science {\bf 258}, 1122 (1992).
\bibitem{perk}
 T.T. Perkins and S.R. Quake and D.E. Smith and S. Chu,Science {\bf 264}, 8222 (1994).

 \bibitem{strick}
T.R. Strick, J.-F. Allemand, D. Bensimon, A. Bensimon and V. Croquette,
Science {\bf 271}, 1835 (1996).

\bibitem{marsig}
C. Bustamante, J.F. Marko, E.D. Siggia and S. Smith, Science {\bf 265},
1599 (1994);
 A. Vologodskii, Macromolecules {\bf 27}, 5623 (1994).

\bibitem{fixman}
M. Fixman and J. Kovac, J. Chem. Phys. {\bf 58}, 1564 (1973).

\bibitem{sDNA}
P. Cluzel, A. Lebrun, C. Heller, R. Lavery, J.L. Viovy,
D. Chatenay and F. Caron, Science {\bf 271}, 792 (1996); S.B. Smith, Y. Cui
and C. Bustamante, Science {\bf 271}, 795 (1996).

\bibitem{strick2}
T.R. Strick, J.-F. Allemand,  D. Bensimon and V. Croquette, Biophys. Jour
{\bf 74} 2016 (1998).

\bibitem{strick3}
T.R.Strick, V. Croquette, and D. Bensimon.
 Proc. Natl. Acad. Sci (USA) {\bf 95 } 10579, (1998).

\bibitem{houch98} T.R. Strick , J.-F. Allemand,  D. Bensimon,
 V. Croquette, C. Bouchiat, M. M\'ezard and R. Lavery,
 Proceedings of August 1998 
les Houches Summer School, to be published.

\bibitem{bouc99}
C.~Bouchiat, M.D. Wang, S.~M. Block, J.-F. Allemand, and V.Croquette;.
Biophys. J. {\bf 76 },409, (1999).

\bibitem{pDNA} 
J.-F. Allemand,  D. Bensimon, R. Lavery and V. Croquette,
 Proc. Natl. Acad. Sci (USA),{\bf 95},14152  (1998).
 
\bibitem{BM1}
C. Bouchiat and M. M\'ezard, Phys. Rev. Lett. {\bf 80}, 1556 (1998).

\bibitem{liverpool}
T.B. Liverpool, R. Golestanian and K. Kremer, Phys.Rev.Lett.
{\bf 80} (1998) 405.

\bibitem{zhou}
Zhou Haijun and Z.C. Ou-Yang, cond-mat/9901321.

\bibitem{cocco}
S. Cocco and R. Monasson {\em Phys. Rev. Lett} {\bf 83}, 5178 (1999)
\bibitem{fain}
B. Fain and J. Rudnick, cond-mat/9903364.

\bibitem{marsig2}
J.F. Marko and  E.D. Siggia, Science {\bf 265}, 506 (1994);
 Phys. Rev. {\bf E52}, 2912 (1995).
 
\bibitem{BDM}
D. Bensimon, D. Dohmi and M. M\'ezard, Europhys. Lett. {\bf 42}, 97 (1998).

\bibitem{marko}
J.F. Marko, Europhys. Lett. {\bf 38},
183 (1997), and Phys. Rev. E {\bf 57}, 2134 (1998).

\bibitem{KLNO}
R.D. Kamien, T.C. Lubensky, P. Nelson and C.S. O'Hern, Europhys. Lett. {\bf 38},
237 (1997).

\bibitem{mornel1}
J.D. Moroz and P. Nelson, Proc. Natl. Acad. Sci USA {\bf 94}, 14418 (1997).

\bibitem{mornel2}
J.D. Moroz and P. Nelson, Macromolecules  {\bf 3},6333 (1998).


\bibitem{geometry}
J.H. White, Amer. J. Math. {\bf 91}, 693 (1969); F.B. Fuller, Proc. Nat.
Acad.
Sci. USA
{\bf 68}, 815 (1971)

\bibitem{fullerform}
 F.B. Fuller, Proc. Nat. Acad. Sci. USA
{\bf 75},3557  (1978)

\bibitem{fain} 
B. Fain,J. Rudbick,and  S. \"{O}stlund 
Phys. Rev. {\bf E55},7364 (1997)

\bibitem{boles}
T.C. Boles, J.H. White and N.R. Cozzarelli, J. Mol. Biol. {\bf 213}, 931
(1990).

\bibitem{adrian}
M. Adrian, B. ten Heggeler-Bordier, W. Whali, A.  Z.Stasiak, A. Stasiak and J. Dubochet.
EMBO J. {\bf 13}, 451

\bibitem{rw}
See the  discussion in A. Comtet, J. Desbois and C. Monthus,
J. Stat. Phys. {\bf 73}, 433 (1993), and references therein, particularly:
F. Spizer, Trans. Am; Math. Soc. {\bf 87}, 187 (1958); C. Belisle, Ann.
Prob. {\bf 17}, 1377 (1989).

\bibitem{volog}
See A.V. Vologodskii, S.D. Levene, K.V. Klenin, M. Frank-Kamenetskii and
N.R. Cozzarelli,
J. Mol. Biol. {\bf 227}, 1224 (1992) and references therein.

\bibitem{marvol}
J.F. Marko and A.V. Vologodskii, Biophys. Journ. {\bf 73}, 123 (1997).
\end{references}
\end{document}